\documentclass{aa}  
\usepackage{subcaption}
\usepackage{mathrsfs}
\usepackage{graphicx}
\usepackage{txfonts}
\newcommand{\hii} {H\,{\sc ii}}

\newcommand{\Teff} {T$_{\rm eff}$}

\newcommand{\Msun} {$M_{\odot}$}
\newcommand{\Rsun} {$R_{\odot}$}
\newcommand{\Lsun} {$L_{\odot}$}
\newcommand{\kms} {km\,s$^{-1}$}
\newcommand{\ioni}[2]{{#1\,\sc{#2}}}
\begin{document}
\title{The massive multiple system HD~64315 \thanks{Based on observations 
obtained at the European Southern Observatory under programmes 078.D-0665(A), 082-D.0136
and 093.A-9001(A).
Based on observations made with the Nordic Optical Telescope, operated on
the island of La Palma jointly by Denmark, Finland, Iceland, Norway, and
Sweden, in the Spanish Observatorio del Roque de los Muchachos of the
Instituto de Astrof\'{\i}sica de Canarias.
}} 
\titlerunning{The multiple system HD\,64315}
%
%\subtitle{}
%
\author{J. Lorenzo\inst{1}
    	\and
	S. Sim\'{o}n-D\'{\i}az\inst{2, 3}
    	\and
	I. Negueruela\inst{1}
    	\and
    	F. Vilardell\inst{4}
    	\and
        M. Garcia\inst{5}
        \and
        C. J. Evans\inst{6}
        \and
        D. Montes\inst{7}
}
\offprints{J. Lorenzo, \email{javihd64315@gmail.com}}
\institute{ Dpto. de F\'{\i}sica, Ingenier\'{\i}a de Sistemas y Teor\'{\i}a de la Se\~nal, Escuela 
            Polit\'ecnica Superior, Universidad de Alicante, Carretera San Vicente del Raspeig s/n,
E03690 San Vicente del Raspeig, Spain
	    \and
            Departamento de Astrof\'{\i}sica, Universidad de La Laguna, E38205 La Laguna, Tenerife, Spain
            \and
            Instituto de Astrof\'{\i}sica de Canarias, E38200 La Laguna, Tenerife, Spain.            
            \and
            Institut d'Estudis Espacials de Catalunya, Edifici Nexus, c/ Capit\'a, 2-4, desp. 201, E08034 Barcelona, Spain
            \and
            Centro de Astrobiolog\'{\i}a (CSIC/INTA), Instituto Nacional de T\'ecnica Aeroespacial, E28850 Torrej\'on de Ardoz, Madrid, Spain
            \and
            UK Astronomy Technology Centre, Royal Observatory Edinburgh, Blackford Hill, Edinburgh EH9 3HJ, UK
            \and
            Departamento de Astrof\'{\i}sica y Ciencias de la Atm\'osfera, Facultad de Ciencias F\'{\i}sicas, Universidad Complutense de Madrid, E28040 Madrid, Spain
           }
\date{Received ...; accepted ...}

% \abstract{}{}{}{}{} 
% 5 {} token are mandatory
\abstract
{The O6\,Vn star HD\,64315 is believed to belong to the star-forming region known as NGC~2467, but previous distance estimates do not support this association. Moreover, it has been identified as a spectroscopic binary, but existing data support contradictory values for its orbital period.} 
{We explore the multiple nature of this star with the aim of determining its distance, and understanding its connection to NGC~2467.}  
{A total of 52 high-resolution spectra have been gathered over a decade. We use their analysis, in combination with  the 
photometric data from {\em All Sky Automated Survey} and Hipparcos catalogues, to conclude that HD\,64315 is composed of 
at least two spectroscopic binaries, one of which is an eclipsing binary. We have developed our own program to fit four 
components to the combined line shapes. Once the four radial velocities were derived, we obtained a model to fit the 
radial-velocity curves using the Spectroscopic Binary Orbit Program (SBOP). We then implemented the radial velocities of the 
eclipsing binary and the light curves in the Wilson-Devinney code iteratively to derive stellar parameters for its components. 
We were also able to analyse the non-eclipsing binary, and to derive minimum masses for its components which dominate the system flux.} 
{HD\,64315 contains two binary systems, one of which is an eclipsing binary. The two binaries are separated by $\sim0.09\:$
arcsec (or $\sim500\:$AU) if the most likely distance to the system, $\sim5\:$kpc, is considered.
The presence of fainter companions is not excluded by current observations. The non-eclipsing binary (HD\,64315\,AaAb) has a period of 
$2.70962901\pm0.00000021\:$d. Its components are hotter than those 
of the eclipsing binary, and dominate the appearance of the system. The eclipsing binary (HD\,64315\,BaBb) has a shorter period of 
$1.0189569\pm0.0000008\:$d. We derive masses of $14.6 \pm 2.3\:$\Msun\ for both components of the BaBb system. They are almost identical; 
both stars are overfilling their respective Roche lobes, and share a common envelope in an overcontact configuration. 
The non-eclipsing binary
is a detached system composed of two stars with spectral types around O6\,V with minimum masses of $10.8\:$\Msun\ and $10.2\:$\Msun, 
and likely masses $\approx30\:$\Msun.} 
{HD\,64315 provides a cautionary tale about high-mass star isolation and multiplicity. Its total mass is likely above $90\:$\Msun, but it seems to have formed without an accompanying cluster. It contains one the most massive overcontact binaries known, a likely merger progenitor in a very wide multiple system.}
\maketitle
%
% ------------------------------------------------------------------------
\section{Introduction}\label{intro}
% ------------------------------------------------------------------------

HD~64315 (HIP~38430, CD~$-26\degr$5115, V402~Pup) is the main ionising source of the
Galactic \hii\ region Sh2-311 \citep{Sha59}. This bright nebulosity (and hence the 
star itself) is apparently connected to a number of dark and bright clouds, 
extending over almost 1$\degr$ on the sky in the region of Puppis, which is 
frequently referred to as NGC~2467. Originally, NGC~2467 was believed to be a 
large cluster, but several authors have concluded that it is actually the 
projection of a number of bright foreground B- and A-type stars on top of a 
distant star-forming association \citep{lode1966, fein1989}. 
Within these clouds, two compact young open clusters, Haffner 18 and Haffner 19, 
were also found to be illuminating smaller H\,{\sc ii} regions. 
Haffner 18 contains an O7\,V star and probably three O9 stars \citep{mo05}. 
Haffner~19 only contains stars up to B0--B1 \citep{mu96}. 
Examination of wide field images of the area very strongly conveys the
impression of a single star-forming region, including the two clusters
and NGC~2467 (now understood only as the area surrounding HD~64315), an idea 
also supported by analysis of {\it Spitzer} observations of the area \citep{sn09}.
Several authors have investigated this hypothesis by deriving distances to the
three clusters, obtaining discrepant results. \cite{fi74} placed Haffner~18 and 
Haffner~19 at 6.9~kpc, a much larger distance than the 3.7\,--\,4.4~kpc calculated 
for HD~64315 and its associated \hii\ region \citep{geor1970, cruz1974, pism1976}. More recent 
studies suggest that Haffner~18 is more distant than Haffner~19 \citep{mu96, mu98, mo02, mo05, yadav15}, 
the latter having a distance more compatible, but still larger than NGC~2467.

In the case of HD\,64315, the few distance determinations found in the literature 
\citep{cruz1974, pism1976} are based on photometry and
the assumption that it is a single star. However, this star, initially classified 
as O6Vn \citep{Wal82}, was found to be a double-lined spectroscopic binary
by \cite{soli1986}. By using medium dispersion spectrograms obtained at 
CTIO between 1982 and 1984, \citeauthor{soli1986} identified the two components
as $\sim$\,O6 stars and derived preliminary orbital elements for the binary system, obtaining a period of 1.34 days and a mass ratio of $\sim$0.83. In view of this binary nature, all the studies mentioned above must have underestimated the distance to the star.

With the aim of investigating the orbital and stellar properties of this binary system, we obtained extensive 
high-resolution, high signal-to-noise ratio (S/N) spectra of HD\,64315 with 
the Fiber Extended Range Optical (FEROS; \cite{kauf99}) and the FIbre-fed Echelle Spectrograph (FIES; \cite{telt2014}) spectrographs. 
In the analysis process, we found strong signatures of 
more than two components present in the spectra, which complicated the 
spectroscopic analysis of the star, but allowed us to discover (in combination
with other photometric and spatial information) that HD\,64315 is in fact 
a multiple system comprising at least four components. 

In this paper, we present observational evidence of the multiple nature of 
HD\,64315 and its implications for the distance
determination to this stellar system. The paper is structured as follows.
The spectroscopic observations and the photometric data are presented in Sect.~\ref{observa}. The spectra and 
their spectral classification are discussed in Sect.~\ref{describe}. A preliminary discussion on the distance 
to the source is made in Sect.~\ref{distance}, based on the interstellar lines present in the spectra.
We then present evidence in Sect.~\ref{evidences} showing that HD\,64315 is composed of two spectroscopic binaries. 
In Sect.~\ref{sb4} we describe the procedure to extract the radial velocities 
of each component from the spectra and develop a comprehensive analysis to obtain the orbital and stellar parameters
for both binary systems. We conclude with the direct estimation of the distance in Sect.~\ref{ded}.
The discussion of results is presented in Sect.~\ref{discusion}, and our main conclusions are presented  
in Sect.~\ref{summary}.

% ------------------------------------------------------------------------
\section{Observations}\label{observa}
% ------------------------------------------------------------------------
A total of 104 spectra grouped in 25 observing blocks (OBs) were obtained in service mode at 
random phases between 2006 October and 2007 March with the FEROS instrument 
at the ESO/MPG\,2.2\,m telescope\footnote{FEROS is a fixed configuration 
instrument (with R\,=\,48000), giving a wide wavelength coverage of 
3600-9200 \AA\ in one exposure.}(see Table~\ref{t1}; spectra numbered from 1 to 25). 
This first set of spectra was complemented with seven further OBs (14 spectra; see Table~\ref{t1}; spectra numbered from 26 to 32) obtained by CJE observations on 5 nights in 2009 March.
This time, they were observed at specific phases, with some of the OBs 
separated by only a few hours.
The third set of spectra was obtained in the framework of the IACOB project \citep{simo2015} with the high-resolution FIbre-fed Echelle Spectrograph (FIES) 
attached to the Nordic Optical Telescope (NOT), located at the Observatorio del Roque de Los Muchachos (La Palma, Spain) 
between 2013 January 29 and 31 (11 spectra; see Table~\ref{t1}; spectra numbered from 33 to 43).
Finally, the last group (see Table~\ref{t1}; nine spectra numbered from 44 to 52)
was also taken with FEROS by DM during a long run in 2014 May.
All the FEROS spectra were reduced using the reduction pipeline that runs under the MIDAS 
environment \citep{kauf99}. The spectra from each observing block
were combined in order to have a higher signal-to-noise ratio (S/N) and 
eliminate possible cosmic ray contamination. Those taken with FIES 
were homogeneously reduced using the
FIEStool\footnote{http://www.not.iac.es/instruments/fies/fiestool/FIEStool.html} software in advanced mode. A complete 
set of bias, flat, and arc frames obtained on each night were used to this end.  For wavelength calibration, we used arc 
spectra of a ThAr lamp.
In most cases 
the S/N of the target spectra is in excess of 70 per resolution element; half of them have a S/N above 100. 
The spectra were normalised and heliocentric corrections were applied using our own code
developed in IDL. The final set of spectra is summarised in Table~\ref{t1}, where we also show
the exposure time and S/N for every spectrum.
%
% -----------------------------------------------------------------------
% --------------------           TABLE 1         ------------------------
% -----------------------------------------------------------------------
%
\begin{table}[!h]
\caption{Log of spectroscopic observations. Spectra are numbered and
sorted according to ascending dates and we have also included the exposure time,
S/N, and the instrument used.\label{t1}}
\centering 
\scalebox{0.85}{
\begin{tabular}{c c c c c c}
\hline \hline
\noalign{\smallskip}
\#&HJD&UT date&Exposure&S/N&Instrument\\
&-2450000&year-month-day&time (s)& \\
\noalign{\smallskip}
\hline
\noalign{\smallskip}  
  1 & 4011.87680  & 2006-10-03 & 4$\times$300	&120 & FEROS\\   
  2 & 4014.83998  & 2006-10-06 & 5$\times$300	& 68 & FEROS\\   
  3 & 4016.82509  & 2006-10-08 & 4$\times$300	&122 & FEROS\\   
  4 & 4020.79661  & 2006-10-12 & 4$\times$300	&123 & FEROS\\   
  5 & 4020.89005  & 2006-10-12 & 4$\times$300	&156 & FEROS\\   
  6 & 4022.87611  & 2006-10-14 & 4$\times$300	& 72 & FEROS\\   
  7 & 4071.80481  & 2006-12-02 & 4$\times$300	&131 & FEROS\\   
  8 & 4072.76861  & 2006-12-03 &  300	& 44 & FEROS\\   
  9 & 4073.80821  & 2006-12-04 & 4$\times$300	&152 & FEROS\\   
 10 & 4074.82240  & 2006-12-05 & 4$\times$300	&158 & FEROS\\   
 11 & 4075.81631  & 2006-12-06 & 4$\times$300	&158 & FEROS\\   
 12 & 4076.78727  & 2006-12-07 & 4$\times$300	&131 & FEROS\\   
 13 & 4077.74138  & 2006-12-08 & 4$\times$300	&128 & FEROS\\   
 14 & 4080.70283  & 2006-12-11 & 4$\times$300	&113 & FEROS\\   
 15 & 4080.81634  & 2006-12-11 & 4$\times$300	&123 & FEROS\\   
 16 & 4083.84377  & 2006-12-14 & 4$\times$300	&130 & FEROS\\   
 17 & 4084.81600  & 2006-12-15 & 4$\times$300	&149 & FEROS\\   
 18 & 4085.79417  & 2006-12-16 & 4$\times$300	&154 & FEROS\\   
 19 & 4086.76633  & 2006-12-17 & 4$\times$300	&141 & FEROS\\   
 20 & 4088.70577  & 2006-12-19 & 4$\times$300	&143 & FEROS\\   
 21 & 4088.84428  & 2006-12-19 & 4$\times$300	&161 & FEROS\\   
 22 & 4090.60271  & 2006-12-21 & 4$\times$300	&125 & FEROS\\   
 23 & 4091.63779  & 2006-12-22 & 4$\times$300	&116 & FEROS\\   
 24 & 4136.71099  & 2007-02-05 & 4$\times$300	&144 & FEROS\\   
 25 & 4209.54994  & 2007-04-19 & 6$\times$300	&119 & FEROS\\   
 26 & 4909.52528  & 2009-03-19 &  600	& 87 & FEROS\\   
 27 & 4909.53300  & 2009-03-19 &  600	&105 & FEROS\\   
 28 & 4911.50573  & 2009-03-21 &  900	&131 & FEROS\\   
 29 & 4911.66461  & 2009-03-21 &  900	&123 & FEROS\\   
 30 & 4912.49930  & 2009-03-21 &  600	& 97 & FEROS\\   
 31 & 4914.49464  & 2009-03-23 &  600	& 87 & FEROS\\   
 32 & 4914.61375  & 2009-03-24 &  600	&101 & FEROS\\   
 33 & 6322.43998  & 2013-01-29 &  900	& 79 & FIES\\   
 34 & 6322.49056  & 2013-01-29 &  900	& 84 & FIES\\   
 35 & 6322.54242  & 2013-01-30 &  900	& 77 & FIES\\   
 36 & 6322.60285  & 2013-01-30 &  900	& 75 & FIES\\   
 37 & 6322.64116  & 2013-01-30 &  900	& 70 & FIES\\   
 38 & 6323.43504  & 2013-01-30 & 1200	& 76 & FIES\\   
 39 & 6323.48378  & 2013-01-30 & 1200	& 71 & FIES\\   
 40 & 6323.51648  & 2013-01-31 & 1200	& 63 & FIES\\   
 41 & 6323.55005  & 2013-01-31 & 1200	& 60 & FIES\\   
 42 & 6323.59720  & 2013-01-31 & 1200	& 56 & FIES\\   
 43 & 6323.61207  & 2013-01-31 & 1200	& 56 & FIES\\   
 44 & 6789.47231  & 2014-05-11 &  900	& 82 & FEROS\\   
 45 & 6789.55915  & 2014-05-12 &  900	& 78 & FEROS\\   
 46 & 6790.48368  & 2014-05-12 &  900	& 86 & FEROS\\   
 47 & 6791.47097  & 2014-05-13 &  900	& 98 & FEROS\\   
 48 & 6792.47602  & 2014-05-14 &  900	& 96 & FEROS\\   
 49 & 6792.53441  & 2014-05-15 &  600	& 77 & FEROS\\   
 50 & 6793.45206  & 2014-05-15 &  300	& 63 & FEROS\\   
 51 & 6794.47959  & 2014-05-16 &  300	& 56 & FEROS\\   
 52 & 6796.48556  & 2014-05-18 &  600	& 82 & FEROS\\   
\noalign{\smallskip}
\hline
\end{tabular}}

\end{table}
% -----------------------------------------------------------------------
% -----------------------------------------------------------------------

\begin{figure}[h]
\centering
\includegraphics[width=8.cm,angle=0, clip]{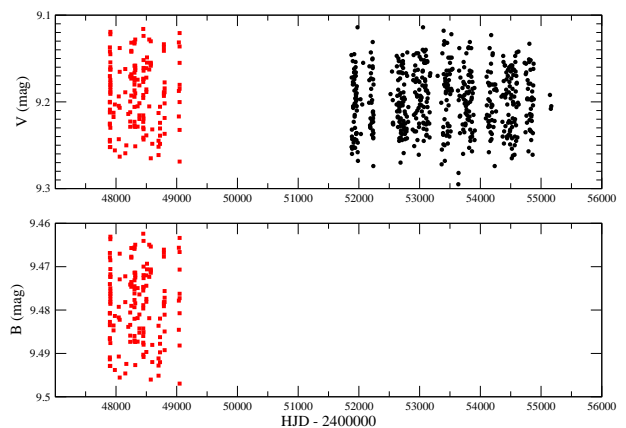}
\caption{Photometric data extracted from the {\em All Sky Automated Survey} (black dots)
and \textit{Hipparcos} catalogues (red dots). The $V$ passband is shown in the top panel. The lower panel shows the $B$-band data.}
\label{f1}
\end{figure}

The spectroscopic observations were complemented with photometric data from
the {\em All Sky Automated Survey}\footnote{www.astrouw.edu.pl/asas} 
\citep[ASAS][]{Poj03} and observations taken by the \textit{Hipparcos} satellite in the $H_p$ band.
There were a total of 544 photometric datapoints
extracted from the {\em All Sky Automated Survey} catalogue (all in the $V$ band). We chose the photometric data corresponding to aperture {\sc MAG\,2},
which has the smallest intrinsic error ($\sigma=$0.034~mag). They are displayed
in Figure~\ref{f1} (top panel). The $H_p$ passband embraces the $V$ and $B$ passbands. The transformations
from $H_p$ to Johnson filters are comprehensively described in \cite{har1998}. The number of photometric points
extracted from the \textit{Hipparcos} catalogue is 149 for every passband (shown in both panels of Figure~\ref{f1}). In total, 
we gathered 693 points in the $V$ filter and 149 in the $B$ filter. We observe the same photometric variability  
in the data from the ASAS and Hipparcos catalogues.

% ------------------------------------------------------------------------
\section{Description of the spectra}\label{describe}
\begin{figure*}[t]
\centering
\includegraphics[width=7cm,angle=90, clip]{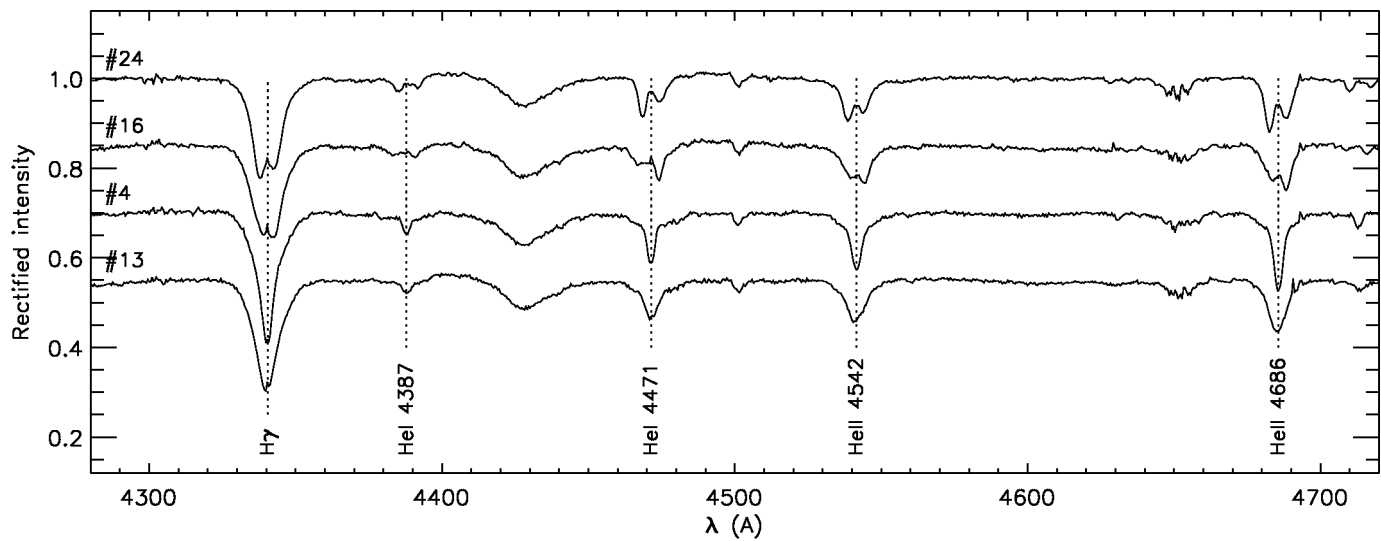}
\caption{Representative examples of the spectrum of HD~64315 at four
different phases. Spectra are numbered for ease of identification.}
\label{f2}
\end{figure*}

% ------------------------------------------------------------------------
Figure \ref{f2} illustrates four representative examples of the spectrum
of HD\,64315 at different phases. We plot the wavelength range between
4300 \AA\ and 4710 \AA, where the main lines used to define the spectral
type (SpT) in mid-O stars are found \citep[see e.g.][]{walb1990}. The first spectrum
from the bottom (\#13) is probably similar to the one analysed by \cite{Wal82}.
Following the Morgan-Keenan (MK) system of spectral classification this spectrum
can be classified as $\sim$O6\,Vn (where the suffix n indicates that the lines are broad).
Continuing to the top of the figure, the third spectrum (\#4) can be also classified
as O6\,V; however this time, the lines are narrower. In addition, the \ion{He}{ii}\,4686 line
is now stronger than the \ion{He}{i}\,4471\,\AA\ and \ion{He}{ii}\,4542\,\AA\ lines. Spectra with 
this type of morphology have been given the `Vz' qualifier 
\cite[see e.g.][]{Wal07}, and hypothesised to correspond to lower (visual) luminosity 
and younger ages \citep[but see discussion by][]{sabi2014}. 
In the two other spectra (\#16 and \#24), it becomes clear that the star is a double-lined 
spectroscopic binary (as seen most clearly in spectrum \#24 in the \ion{He}{i}\,4471\,\AA\ line). 
Previous spectral classifications are thus the consequence of the morphological analysis of spectra 
obtained at phases in which the two components are blended together. 

More recently, \cite{ari2016}, accepting the multiplicity of the system, classified HD\,64315 as O5.5\,V+O7\,V, 
removing the system from the `Vz' category.
At classification resolution, only two components are identified. 
From a first inspection of some of the spectra with maximum separation between both
components (see e.g.\ spectra \#16 and \#24 in Figure \ref{f2}), the spectral types could 
be estimated as O6 and $\sim$\,O5\,--\,O5.5 for the components with stronger and fainter 
lines, respectively. Curiously, the component with fainter lines seems to have an
earlier spectral type. In addition, the broadening of the lines from this component 
is noticeably different in some of the spectra with a similar separation between the
two components.
Indeed, a closer inspection of 
the global spectrum at certain phases already gives strong hints of the presence of more than two components, 
all moving rather quickly in radial velocity.

% ------------------------------ GRAPH 1 ----------------------------------
% -------------------------------------------------------------------------
%

\section{Distance determination using interstellar lines}\label{distance}
\begin{figure}[h]
\centering
\includegraphics[width=8 cm,angle=0,clip]{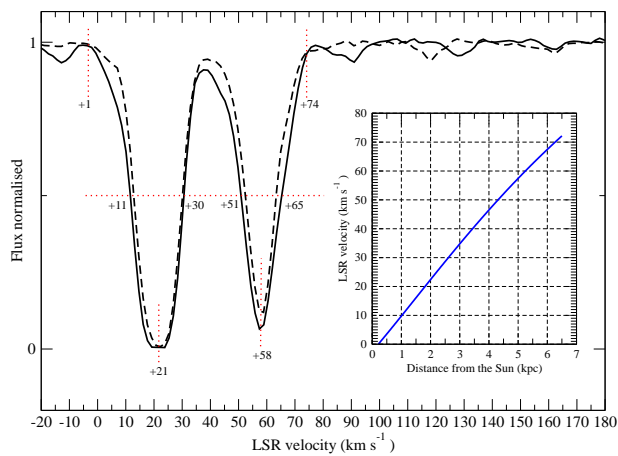}

\caption{Interstellar lines in the spectrum of HD~64315. The main panel shows 
the components of the Na\,{\sc i} D doublet (5889.95\,\AA$\:$ solid line; 5895.92\,\AA$\:$ 
dashed line) in velocity space. The inset shows the Galactic 
rotation curve along this line of sight.  The velocities are measured with respect to the 
local standard of rest (LSR) assuming a solar motion of $+16.6\:{\rm km}\,{\rm s}^{-1}$ 
towards Galactic coordinates $\ell=53\degr$; $b=+25\degr$.}
\label{f10}
\end{figure}

Before presenting the observational evidence for more than two components in HD\,64315
 and its physical characterisation using the available spectroscopic and photometric datasets,
we discuss the kinematic distance to HD\,64315.
Using the interstellar Na\,{\sc i} D lines (5890.0\AA, 5895.9\AA), we studied the radial velocity 
distribution of the interstellar material in the direction towards HD\,64315 
($l=243\fdg15$; $b=+0\fdg36$). We calculated the velocity 
scale with respect to the local standard of rest (LSR) by assuming that the Sun's motion 
with respect to the LSR corresponds to +16.6 \kms\ towards Galactic 
coordinates $l=53\,^{\degr}$; $b=+25\,^{\degr}$. The interstellar Na\,{\sc i} D lines are shown in Figure~\ref{f10} as normalised flux as a function of LSR velocities. 
The two lines have a very similar shape, showing two distinct components. 
Both components display only positive velocities from +1 to $+74\:$\kms. 
None of the components is saturated, and so we can determine their centres. 
The broader feature is centred at $+21\:$\kms, while the narrower feature 
is at $+58\:$\kms.

There are few stars along similar lines of sight with well-studied interstellar lines. HD\,68761, 
with galactic coordinates $l=254\fdg37$; $b=-1\fdg61$, displays only 
one component in its interstellar lines, centred at $+8\:$\kms. This star is 
situated at a distance $\la1.5\:$kpc \citep{hun2006}.
HD\,58978 ($l=237\fdg41$; $b=-2\fdg99$), has several distance estimates, 
ranging from $\sim800\:$pc to $\sim1.7\:$kpc \citep{hun2006}. Its interstellar lines display a main 
component centred at $+11\:$\kms\ and a very weak component with an edge velocity 
around $+29\:$\kms. 

\citet{gy02} measured the velocities of all catalogued molecular clouds in this area, finding that 
all had velocities between +20 and +25 \kms and typical distances $\la1$~kpc. 
All this suggests that the main component in the interstellar lines of HD\,64315 arises from relatively 
nearby clouds. The clear separation of the second component suggests that the extinction is very low at 
intermediate distances. The second component shows velocities corresponding to much higher distances.

\citet{pism1976} found kinematic distances to the \ion{H}{ii} regions around HD\,64315, Haffner 18 
and Haffner 19. Their observations were re-reduced by \citet{mo02}, who find LSR values of +50, +58 and 
+50 \kms, respectively. However, \citet{pism1976} warn that the \ion{H}{ii} velocity 
close to HD\,64315 seems to be affected by the expansion of gas around the ionising star, with smaller 
values of $v_{{\rm LSR}}$ in the immediate vicinity of the star. This interpretation is borne out by 
the presence of interstellar material with higher $v_{{\rm LSR}}$ along the line of sight. The data 
available can be interpreted as suggesting that the actual radial velocity of the complex is at 
least $+58\:{\rm km}\,{\rm s}^{-1}$, with the lower values measured at some points due to expanding 
shells around the ionising stars. This agrees very well with the shape of the weaker component of the interstellar 
lines, centred at $v_{{\rm LSR}}\approx+58\,$\kms. The inset in Figure~\ref{f10} shows the Galactic rotation curve in this direction, 
computed considering 
circular galactic rotation and adopting the rotation curve of \citet{rei14}. 
Along the line of sight towards HD~64315, all radial velocity curves display only positive and 
monotonically increasing values, from a distance around $0.25\:$kpc. 
Assuming $v_{{\rm LSR}}\approx+58\,$\kms, the kinematic distance estimate to HD\,64315
corresponds to $d\approx5\:$kpc, which we adopt as a preliminary distance.

% ------------------------------------------------------------------------
\section{Evidence for more than two components}\label{evidences}
% ------------------------------------------------------------------------

\subsection{Period determination}\label{period}

\begin{figure}[!h]
\centering
\includegraphics[width=8.cm,angle=0,clip]
{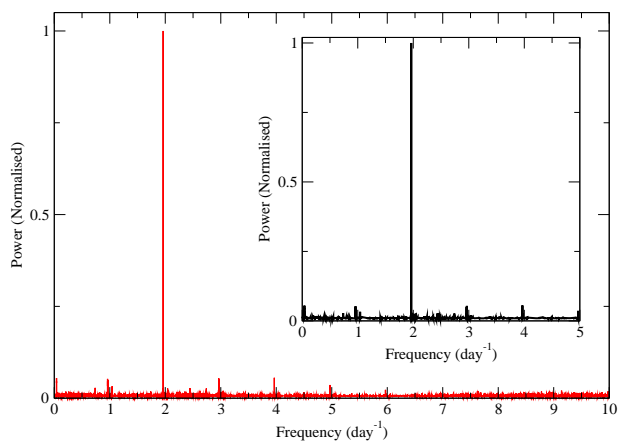}
\caption{Results of the Lomb-Scargle periodogram for the photometric data in the $V$ filter (left). 
The result of applying the {\sc clean} algorithm is shown (right). The frequency peak corresponds
to half the orbital period.}
\label{f3}
\end{figure}

We carried out a timing analysis of the photometric data described in Sect.~\ref{observa}.
We used the {\sc period} program inside the {\it Starlink} suite for every passband.
The Lomb-Scargle algorithm \citep{lomb1976,scar1982}, in a range of frequencies 0\,--\,35.7\,d$^{-1}$, gives a photometric period of 
$1.0189650\pm0.000112$~d and $1.018958\pm0.000018$~d for the $B$- and $V$-filter  data, respectively.
The agreement between the two periods is excellent. These values are
confirmed with the {\sc clean} algorithm \citep{robe1987}, which removes spurious periods caused by the 
window function. The Lomb-Scargle periodogram in a range of frequencies 0\,--\,10\,d$^{-1}$ and an inset of the 
{\sc clean} periodogram, up to $\nu= 5\:$d$^{-1}$ are shown in
Figure~\ref{f3}. There is clearly only one peak above significance. 

\begin{figure}[h]
\centering
\includegraphics[width=8.cm,angle=0,clip]
{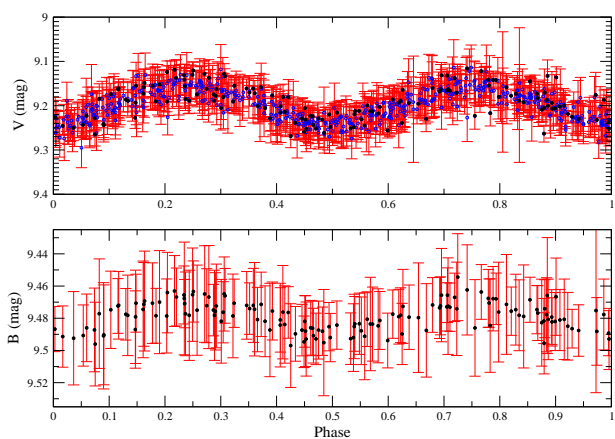}
\caption{Photometric light curves ($V$ filter, top; $B$ filter, bottom), including data and the corresponding error bars from
ASAS (blue open dots) and {\it Hipparcos} (black solid dots).}
\label{f4}
\end{figure}

Figure\ref{f4} shows all the photometric data for both passbands folded on the derived period. The corresponding 
light curves in the $V$ and $B$ filters show an amplitude of $\approx0.18$~mag and $\approx0.05$~mag, respectively. Error bars of the data are also displayed
in Figure~\ref{f4}, showing that most of the errors exceed the amplitude of the light curves
(probably due to the process of transformation to the standard system). The shallow amplitudes and short-period observed might suggest that 
the variability is due to ellipsoidal light variations \citep{wils1976b}. However, as we shall see, the 
light from the system is not dominated by the binary producing the photometric variability, and the variations represent an eclipse.

\subsection{Spectroscopy and radial velocities}

\cite{soli1986} determined a spectroscopic
period of 1.34\,d, based on around 20 medium dispersion spectrograms, presenting a large dispersion
in radial velocities with respect to the model curve fitted. The difficulty in obtaining a good fit is evidenced 
by a difference of 40~\kms\ between the systemic velocities of the two components.
A more recent attempt to characterise HD\,64315 as an eclipsing binary was carried out by our group \citep{lore2010}. We used the first 24 spectra
described in Table~\ref{t1} and obtained a binary period of 2.71~d (twice the period of \citealt{soli1986}), with the same systemic velocity for the two stars. Even so, the radial velocity curves showed an unacceptable standard deviation of 35\,\kms; the residuals of the radial velocities increase at phases around
zero and 0.5 (i.e.\ near the eclipses). Given the total disagreement between the photometric and spectroscopic period and the hints of the presence of more stellar components, we decided to schedule further observations so that some spectra were taken at the same phases with respect to the photometric period (1.01896~d), while other spectra were taken separated by just a couple of hours, a very small phase shift with respect to the spectroscopic period. With these constraints, we expected to reveal changes in the morphology (line profiles), while keeping
the components of the eclipsing binary at the same radial velocity.

\begin{figure}[h]
\centering
\includegraphics[width=8. cm,angle=0,clip]
{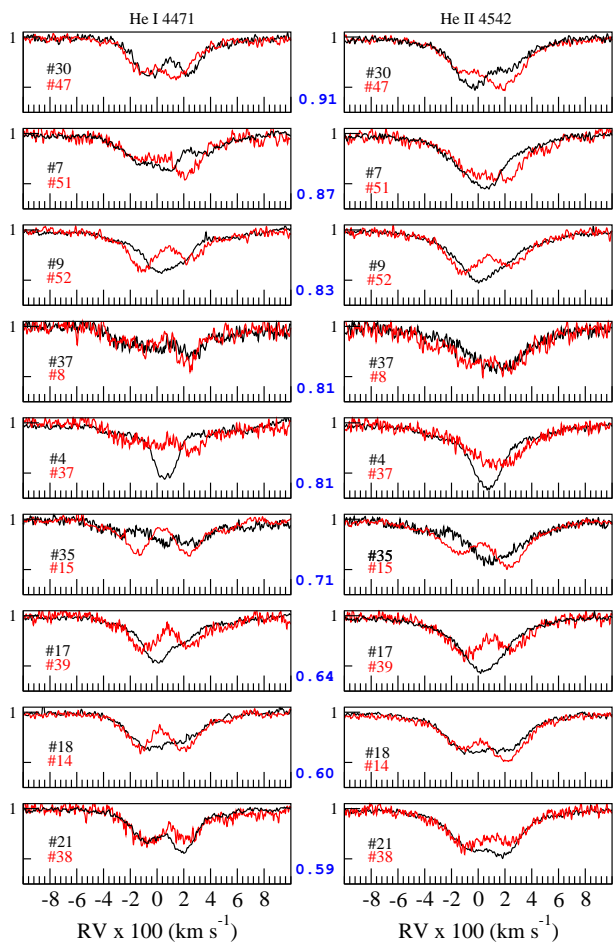}
\caption{Comparison of several pairs of spectra taken on different nights, but corresponding to the same phase (blue numbers) 
of the 1.019~d photometric period. The presence of more than two components is obvious.}
\label{f5}
\end{figure}

As a consequence of this successful strategy, observations between 2009 March and 2014 May have been much more useful to 
understand how many components the system has and how each one of contributes to the combined spectrum. As examples, 
in Figure~\ref{f5}, we show a set of pairs of spectra, represented by the \ion{He}{i}\,4471 and \ion{He}{ii}\,4542 lines, 
which define spectral type for O-type stars. Every pair was observed at a very similar phase ($\phi$), according to
the photometric period, but they were obtained on a different date. For instance, the spectra 
\#30 and \#47 (top panel of Figure~\ref{f5}) were acquired with a time difference of more than 5 years, but
at phase $\phi \approx0.91$ (according the period of the eclipsing binary, 1.019 d). First of all, they do not show the
same morphology. Moreover, the line 
morphology is not typical of a binary close to eclipse, where the radial velocity of
both components should be close to systemic velocity. 
Rather \ion{He}{ii}\,4542 shows a double line in spectrum \#30. A comparable situation happens for the pairs of spectra  
\#7, \#51 ($\phi$=0.87) and \#9, \#52 ($\phi$=0.83). 
However, the spectra \#37, \#8 ($\phi$=0.81) have a time difference of more than 6 years, but 
their profiles are morphologically very similar. 
 
We display a few more examples in Figure~\ref{f5} to emphasise the complexity of this stellar system. The only sensible conclusion 
after all these observations is that the stars forming the spectroscopic binary (or, at least dominating the spectrum) are not 
the same ones giving the photometric signal at 
1.019~d. The 2.71~d period found from the radial velocity analysis \citep{lore2010} corresponds to a different binary.

Once we have established that there is a non-eclipsing SB2 binary giving the prevailing spectroscopic period of 2.71~days and an eclipsing binary causing the photometric 1.019~d signal, we can guess that the components of the SB2 system should be more luminous than the stars in the eclipsing binary system, as they dominate the combined flux. Since two stars with spectral types not very far from O6 are seen, the spectral types of the components of the eclipsing binary must be later than O7\,V, so that they are later than the secondary star in the non-eclipsing binary system.

\subsection{Interferometry and parallax}

Using \textit{Speckle} interferometry obtained over a period of a decade, 
\citet{maso2009} derived the angular separation of HD\,64315, resolving at least two components 
with a separation of $0\farcs091$. In Sect.~\ref{distance}, we discussed the distance to HD~64315
and estimated $d\approx5\:$kpc. At this distance, the geometric distance between the components would be $455\:$AU 
($\approx100\,000\:$\Rsun). This separation is certainly not consistent with the short orbital periods that we have found. 
Furthermore, \citet{toko2010} obtained \textit{Speckle} interferometry images of HD~64315 (see Figure\,12 in the mentioned study) suggesting 
that the system may consist of three visual components
in a linear configuration \citep{toko2010}. The fit  of  the $x$-axis  scan  with  two components is marginal, 
but experiments with three component made little or no improvement \citep{aldo2015}. Therefore the two visual 
components resolved represent two systems of comparable (but not equal) brightness that are either in a very wide orbit or are not bound. 
The simplest possibility is that each the visual components represents one of the two binaries that we have identified 
(one in the photometric signal and the other in the spectra). Hereafter, we name the non-eclipsing binary system 
A, composed of stars Aa and Ab, while the eclipsing binary is known as B, with components Ba and Bb. Both  
are spectroscopic binaries (SB2), but the lines corresponding to Ba and Bb are mostly hidden within the complex and broad 
profiles generated by the more luminous components of system A. In the following section (Sect.~\ref{sb4}), we describe the 
technique developed to disentangle the spectra and the subsequent orbital analysis.

% ------------------------------------------------------------------------
\section{HD\,64315, a double spectroscopic binary}\label{sb4}

\subsection{Determination of radial velocities } \label{rvdet}

In Sect.~\ref{evidences}, we presented inescapable evidence that
HD~64315 is a double spectroscopic binary, with the non-eclipsing binary (nEB = A) composed of two mid-O
stars and the eclipsing binary (EB = B) containing two stars of lower temperature. The orbital period
of B is obviously the photometric period derived in Sect.~ \ref{period}, while the orbital period of A must be around 2.71~d. Our spectroscopic monitoring (see Table~\ref{t1}) covers almost 8 years, during which time A has completed more than 1\,000 orbital cycles.

Every one of the four components contributes to every spectrum with its flux. Since we see them in the \ion{He}{i}, \ion{He}{ii} and Balmer lines, all components are O-type stars. According to the classical criteria 
for O-type spectral classification by \citet{walb1990}, the ratio between the \ioni{He}{i}~4471\,\AA\
and \ioni{He}{ii}~4542\AA\ lines is sensitive to temperature and thus we chose these lines to fit
the spectral line shape with multiple functions. The spectral line of a single star can be approximated with a Gaussian function, characterised by three parameters:
amplitude, width and centroid of the peak. Even though line shapes in isolated stars present more complex profiles, the many added complications that are discussed below call for the choice of a simple Gaussian shape. The amplitude and width are dependent on the spectral line, and the centroid is dependent on the time when the spectrum was taken (i.e.\ the radial velocity of the component at that time). Since we have analysed two spectral lines in 52 spectra and we require four Gaussian functions to fit each line profile, our model must have eight widths, eight amplitudes, and 52 positions of the centre of the peaks. Our unknowns are ten orbital parameters:
period, zero point of ephemeris, systemic velocity, and semi-amplitude of the velocity curve for each binary 
($P_{\textrm{A}}$, $T_{0,\textrm{A}}$, $v_{0,\textrm{A}}$, $K_{\textrm{Aa}}$, $K_{\textrm{Ab}}$, $P_{\textrm{B}}$, $T_{0,\textrm{B}}$, $v_{0,\textrm{B}}$, $K_{\textrm{Ba}}$, $K_{\textrm{Bb}}$). 
From the light curves, two of these parameters, $P_{\textrm{B}}$ and $T_{0,\textrm{B}}$, are well determined. This means that we know the phases corresponding to the orbital period of EB for every spectrum. Moreover, given the short orbital periods and large stellar sizes, we can assume that both binary systems are circularised and synchronised, and so
both eccentricities are zero. To derive the radial velocities, we solve the inverse problem:
we vary the free orbital parameters until the sum of four Gaussian functions matches the shape of 
the line observed. The minimization function is constrained by the expression for the radial velocity due to the 
orbital motion derived from the Kepler laws (see Eq.~2.45 in \citealt{hild2001}).
From an analytical point of view, if we consider that $f_{i\alpha}(v)$ is the normalised flux of a 
spectral line $i$ in the spectrum numbered $\alpha$, with a dependence on velocity $v$, and $g_{i\alpha X}(v)$ is
the Gaussian function which represents the flux distribution for star $X$ (Aa, Ab, Ba, Bb)
of the spectral line $i$ (where $i$ can be one of two spectral lines; \ioni{He}{i}\,4471\AA\,
or \ioni{He}{ii}\,4542\AA) in the spectrum $\alpha$, and finally  $M_{i\alpha}(v)$
is the sum of the four Gaussian functions, representing the overall line, given by
 
\begin{equation}
M_{i\alpha}(v)=\sum_{X=1}^{4} (1-g_{i\alpha X}(v))\, ,
\end{equation} 

the function to minimise $F(v)$ would be equivalent to

\begin{equation}
F (v)=\sum\limits_{\alpha, i} |f_{i\alpha}(v)-M_{i\alpha}(v)|
\end{equation}

The function $F(v)$ was implemented via the Python interface called {\em lmfit}. This package
builds complex fitting models for non-linear least-squares problems. Spectra were 
transformed to the velocity space, in a range from $-600\:$\kms
to $+500\:$\kms, and rebinned to 200 bins. The remaining free orbital parameters, $P_{\textrm{A}}$, $T_{0,\textrm{A}}$, 
$v_{0,\textrm{A}}$, $K_{\textrm{Aa}}$, $K_{\textrm{Ab}}$, $v_{0,\textrm{B}}$, $K_{\textrm{Ba}}$ and $K_{\textrm{Bb}}$, were bounded. The sixteen free parameters
to characterise the widths and heights of every Gaussian function were constrained using our knowledge of the system: given that $P_{\textrm{A}}$
is longer than $P_{\textrm{B}}$ and the orbits are synchronised, the rotational velocities of Ba and Bb will be higher than the rotational velocities
of Aa and Ab, and so we expect that Gaussians corresponding to Ba and Bb are wider than those for Aa and Ab.
We consider that the flux of the stars is constant during an orbital cycle and that the stars do not pulsate. Finally, we do not take into account the 
Struve-Sahade \citep{stru1958} and Rossiter-Mclaughlin \citep{ross24, mcla24}  effects.

\begin{figure*}[t]
\includegraphics[width=19cm,height=10cm,angle=0]
{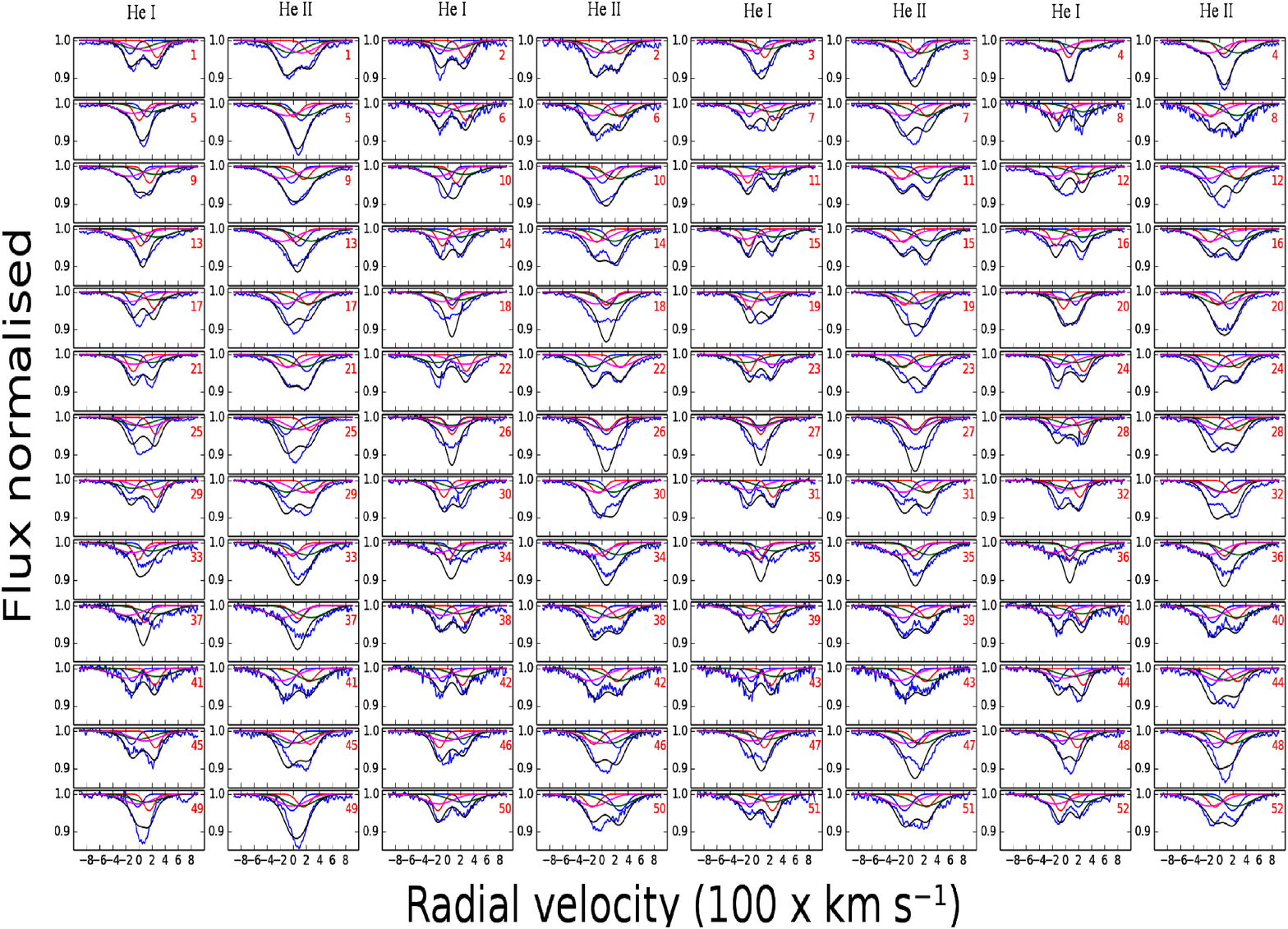}
%{fit52_1.eps}

\caption{Final fit: the observed lines (blue), the sum of four Gaussian functions (black), four Gaussian functions 
in every spectral line (Aa:blue; Ab:red; Ba:green; Bb:magenta). Numbers represent the order of the spectrum (for every
pair: left \ioni{He}{i} and right \ioni{He}{ii})}
\label{f6}
\end{figure*}

The resulting models (104 in total = 2 spectral lines x 52 spectra) are displayed in Figure~\ref{f6}. In each panel, 
we can see the four Gaussian functions, representing the contribution of every component to the spectral line.  
The positions of the peaks are the radial velocities of the different components. To evaluate the goodness of the fit, we present the Pearson product-moment correlation coefficient $R^{2}$
for every fit in Table~\ref{t2}. The coefficient $R^{2}$ is $>0.9$ in $\approx42\%$ of the models
for the \ioni{He}{i} line, and just 8\% of them have $R^{2}$<0.7. In the case of the \ioni{He}{ii} line, $\approx62\%$ of the models have $R^{2}$\,>\,0.9 and only
2\% present $R^{2}$\,<\,0.7. We did not find any correlation between the best fits and the orbital phases, or the observing campaigns. Radial velocities are shown in Table~\ref{t3}. In the next section, we  use
these four sets of radial velocities to derive orbital and stellar parameters.

\begin{table}[!h]
\caption{Pearson product-moment correlation coefficient $R^{2}$ for every line and spectrum. \label{t2}}
\centering 
\scalebox{0.9}{
\begin{tabular}{c c c c c c}
\hline \hline
\noalign{\smallskip}
$\sharp$&$R^{2}$ (\ioni{He}{i})&$R^{2}$ (\ioni{He}{ii})&$\sharp$&$R^{2}$ (\ioni{He}{i})&$R^{2}$ (\ioni{He}{ii})\\
\noalign{\smallskip}
\hline
\noalign{\smallskip}  
     1	& 0.9658 & 0.9732  &	 27  & 0.9161 & 0.8906 \\
     2	& 0.9080 & 0.9758  &	 28  & 0.8847 & 0.8661 \\
     3	& 0.9622 & 0.9710  &	 29  & 0.9072 & 0.8784 \\
     4	& 0.9908 & 0.9947  &	 30  & 0.9278 & 0.9373 \\
     5	& 0.9853 & 0.9854  &	 31  & 0.8759 & 0.8660 \\
     6	& 0.8761 & 0.8916  &	 32  & 0.9421 & 0.8825 \\
     7	& 0.7351 & 0.8326  &	 33  & 0.8006 & 0.9583 \\
     8	& 0.7357 & 0.8810  &	 34  & 0.7078 & 0.9559 \\
     9	& 0.9803 & 0.9895  &	 35  & 0.6737 & 0.9359 \\
    10	& 0.8315 & 0.9744  &	 36  & 0.6231 & 0.9060 \\
    11	& 0.9258 & 0.9736  &	 37  & 0.5603 & 0.8708 \\
    12	& 0.5886 & 0.6960  &	 38  & 0.9376 & 0.9827 \\
    13	& 0.9638 & 0.9666  &	 39  & 0.9507 & 0.9764 \\
    14	& 0.8986 & 0.9369  &	 40  & 0.9113 & 0.9535 \\
    15	& 0.8891 & 0.9649  &	 41  & 0.8898 & 0.9421 \\
    16	& 0.7631 & 0.9373  &	 42  & 0.8169 & 0.9231 \\
    17	& 0.8054 & 0.8816  &	 43  & 0.7397 & 0.9035 \\
    18	& 0.8165 & 0.8603  &	 44  & 0.8923 & 0.8645 \\
    19	& 0.8498 & 0.9027  &	 45  & 0.9299 & 0.8972 \\
    20	& 0.9868 & 0.9903  &	 46  & 0.9155 & 0.8770 \\
    21	& 0.9453 & 0.9962  &	 47  & 0.8925 & 0.8754 \\
    22	& 0.8163 & 0.9823  &	 48  & 0.9089 & 0.9248 \\
    23	& 0.8849 & 0.8668  &	 49  & 0.9488 & 0.9575 \\
    24	& 0.9044 & 0.9746  &	 50  & 0.9255 & 0.8428 \\
    25	& 0.7900 & 0.8872  &	 51  & 0.9279 & 0.9180 \\
    26	& 0.8968 & 0.8978  &	 52  & 0.9147 & 0.9147 \\

\noalign{\smallskip}
\hline
\end{tabular}}
\end{table}

\begin{table}[!ht]
\caption{Radial velocities of HD~64315 derived from the fitting of the
four Gaussian functions in both spectral lines (\ioni{He}{i}\,4471\AA\
and \ioni{He}{ii}\,4542\AA) for every spectrum. \label{t3}}
\centering 
\scalebox{0.8}{
\begin{tabular}{c c c c c c c c c c}
\hline \hline
\noalign{\smallskip}
$\sharp$&RV$_{Aa}$&RV$_{Ab}$&RV$_{Ba}$&RV$_{Bb}$&$\sharp$&RV$_{Aa}$&RV$_{Ab}$&RV$_{Ba}$&RV$_{Bb}$\\
\noalign{\smallskip}
\hline
\noalign{\smallskip}  
{\bf 1 }   &  -130 &   283	 &  -29  &   126 & {\bf  27}    &   73	&   67       &  47   &   53   \\ 	  
{\bf 2 }   &  -118 &   271	 &  109  &   -7  & {\bf  28}    &   -132 &   286      &  142  &   -39  \\ 	  
{\bf 3 }   &  14   &   130	 &  181  &   -77 & {\bf  29}    &   -128 &   281      &  -84  &   180  \\ 	  
{\bf 4 }   &  83   &   56	 &  276  &   -170& {\bf  30}    &   187  &   -55      &  176  &   -72  \\ 	  
{\bf 5 }   &  127  &   10	 &  194  &   -90 & {\bf  31}    &   -106 &   258      &  225  &   -120 \\ 	  
{\bf 6 }   &  -131 &   284	 &  248  &   -142& {\bf  32}    &   -72  &   222      &  68   &   32   \\ 	  
{\bf 7 }   &  -105 &   257	 &  231  &   -125& {\bf  33}    &   143  &   -8       &  206  &   -101 \\ 	  
{\bf 8 }   &  260  &   -133	 &  274  &   -167& {\bf  34}    &   120  &   16       &  255  &   -149 \\ 	  
{\bf 9 }   &  -22  &   168	 &  260  &   -154& {\bf  35}    &   96	&   42       &  285  &   -178 \\ 	  
{\bf 10}   &  6   &   139	&  264  &   -158&  {\bf  36}    &    68	&   72       &  289  &   -182 \\  
{\bf 11}   &  257  &   -129	 &  278  &   -172& {\bf  37}    &   50	&   91       &  275  &   -169 \\ 	  
{\bf 12}   &  -111 &   263	 &  291  &   -184& {\bf  38}    &   -120 &   273      &  177  &   -74  \\ 	  
{\bf 13}   &  103  &   35	 &  275  &   -169& {\bf  39}    &   -110 &   263      &  232  &   -127 \\ 	  
{\bf 14}   &  208  &   -78	 &  189  &   -85 & {\bf  40}    &   -103 &   255      &  260  &   -154 \\ 	  
{\bf 15}   &  243  &   -114	 &  284  &   -177& {\bf  41}    &   -94  &   245      &  280  &   -174 \\ 	  
{\bf 16}   &  270  &   -144	 &  269  &   -163& {\bf  42}    &   -80  &   230      &  291  &   -185 \\ 	  
{\bf 17}   &  -84  &   235	 &  231  &   -125& {\bf  43}    &   -75  &   225      &  291  &   -184 \\ 	  
{\bf 18}   &  67   &   73	 &  185  &   -81 & {\bf  44}    &   -123 &   276      &  122  &   -20  \\ 	  
{\bf 19}   &  230  &   -100	 &  122  &   -20 & {\bf  45}    &   -106 &   258      &  -5   &   104  \\ 	  
{\bf 20}   &  160  &   -26	 &  -22  &   120 & {\bf  46}    &   251  &   -124     &  133  &   -31  \\ 	  
{\bf 21}   &  213  &   -82	 &  176  &   -73 & {\bf  47}    &   20	&   123      &  175  &   -72  \\ 	  
{\bf 22}   &  -132 &   285	 &  -172 &   266 & {\bf  48}    &   -39  &   186      &  193  &   -89  \\ 	  
{\bf 23}   &  238  &   -110	 &  -162 &   256 & {\bf  49}    &   -15  &   160      &  115  &   -13  \\ 	  
{\bf 24}   &  -128 &   281	 &  144  &   -41 & {\bf  50}    &   272  &   -146     &  239  &   -133 \\ 	  
{\bf 25}   &  -109 &   261	 &  -20  &   118 & {\bf  51}    &   -95  &   246      &  230  &   -125 \\ 	  
{\bf 26}   &  69   &   71	 &  59   &   41  & {\bf  52}    &   200  &   -69      &  258  &   -152 \\ 	  
 
\noalign{\smallskip}
\hline
\end{tabular}}
\end{table}

\subsection{Orbital analysis} \label{orb_a}

\subsubsection{Non-eclipsing binary (AaAb)} \label{nEB}

\begin{figure}[h]
\centering
\includegraphics[width=8 cm,angle=0, clip]
{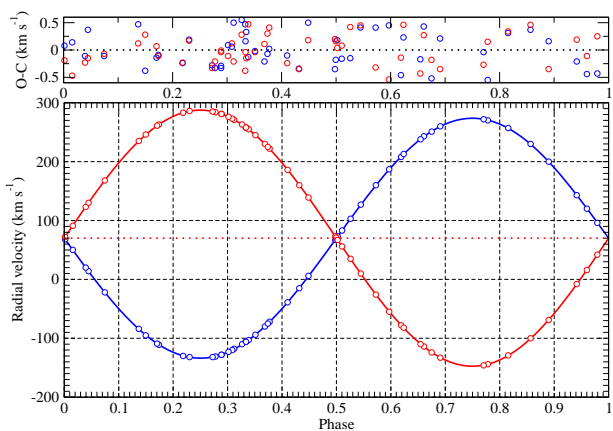}
\caption{Radial velocity curves for the non-eclipsing binary (A), fitted to the observational data and shown against orbital phase 
(blue line: star Aa; red line: star Ab).
The red dotted line corresponds to the systemic velocity.
The residuals are shown in the top panel.}
\label{f7}
\end{figure}

We determined the orbital parameters from the radial velocities derived (Figure~\ref{f6}). We used the Spectroscopic Binary Orbit Program ({\sc SBOP}) by \citet{etze2004}. The guess parameters
are those obtained by the fit described in Sect.~\ref{rvdet}. The radial velocity curve and residuals are displayed in Figure~\ref{f7}. The orbit is circular, as this was an initial assumption. In the case of system A,
we cannot determine the absolute parameters, as we do not know the inclination of the system with respect to the line of sight, but we can obtain a minimum mass for every component. The spectroscopic period determined is very similar to that obtained in \cite{lore2010}, but now the uncertainty is very small, and the residuals are very low, less than $1\:$\kms. These are not the real uncertainties that can be derived from the observations, because, as explained in Sect.~\ref{rvdet}, we imposed a constraint on the model to determine the radial velocities. Radial velocities ordered
by phase are shown in Table~\ref{t3ab}. The parameters derived are shown in
Table~\ref{t5}. The mass ratio is 0.9366, very close to unity. The systemic velocity is the same for both radial velocity curves.  When transformed to the LSR, it is $+55\:$\kms, in very good agreement with the velocity of the \ion{H}{ii} nebula within which HD~64315 is embedded (Sect.~\ref{distance}).

\begin{figure}[h]
\centering
\includegraphics[width=8 cm,angle=0]
{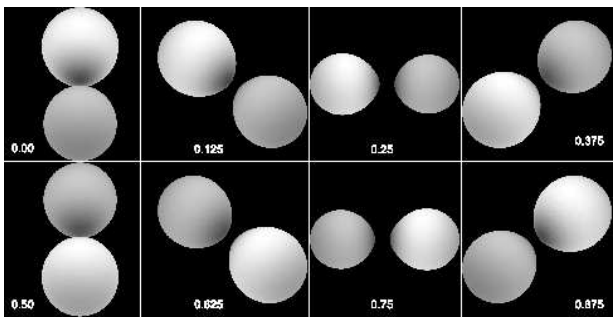}
\caption{Representative drawing of HD\,64315\,AaAb at different phases created with the {\sc phoebe} 
2.0-alpha code via the Python interface. The secondary is slightly greyer and smaller to differentiate it from the primary.}
\label{i4}
\end{figure}

Recently, \citet{ari2016} classified HD~64315 as O5.5\,V +O7\,V. This classification does not take into account the contribution to the spectrum of the eclipsing binary (Ba and Bb). Our results suggest that Aa and Ab are indeed quite similar in mass and temperature. If we calculate the sizes of their Roche lobes, following  
\cite{eggl1983}, we find $R_{\textrm{Aa}}^{\textrm{lobe}}=11.5\:$\Rsun\, and $R_{\textrm{Ab}}^{\textrm{lobe}}=11.2\:$\Rsun. We can estimate the projected rotational velocity by assuming that the two stars just fill their lobe radii. In this case, 
it would be $\sim150\:$\kms\ for both components. As the stars may not fill their Roche lobes, their actual rotational velocity will be higher, in agreement with the expectation of synchronisation. The lack of eclipse implies a lower limit on the orbital inclination. If the stars fill their Roche lobes, it will be around $45\degr$. For this inclination, the masses of the two components are $31\:$\Msun\, and $29\:$\Msun. 
As these masses are consistent with calibrations for a spectral type O6\,V \citep{mart2005}, we do not expect the inclination to be much lower. This strongly suggests that all the assumptions are approximately correct. The semi-major axis of the binary system would then be $32\:$\Rsun.  Given these orbital parameters, this system is
not a contact binary (see Figure~\ref{i4}), but a detached binary system. However, given estimates for synchronisation times in stars with radiative envelopes \citep{zahn75,claret97}, the assumption of synchronisation is fully justified, and so our assumption that the lines of both components of A are narrower than those of Ba and Bb is correct. As mentioned in Sect.~\ref{describe}, the component with fainter lines seems to have an earlier spectral type in some spectra. With our solution, this can be understood as follows: the components both have approximately the same spectral type, around O6\,V, but they appear slightly different in different spectra because changes in the width and height of spectral lines are caused by the many confusing effects inherent to this kind
of early-type close binary (such as Struve-Sahade or tidal distortion) with the added complication of the contribution of the eclipsing binary sibling.

 \begin{table}[h]
\caption{Radial velocities in phase of the non-eclipsing binary (A). \label{t3ab}}
\centering 
\scalebox{0.94}{
\begin{tabular}{c c c c c c c c}
\hline \hline
\noalign{\smallskip}
$\sharp$&phase &RV$_{Aa}$&RV$_{Ab}$&$\sharp$&phase &RV$_{Aa}$&RV$_{Ab}$\\
\noalign{\smallskip}
\hline
\noalign{\smallskip}  
{\bf 36}	&  0.0016 &  68   &   72      &       {\bf 32}        &  0.3772 &  -72  &   222     \\
{\bf 37}	&  0.0157 &  50   &   91      &       {\bf 48}        &  0.4102 &  -39  &   186     \\
{\bf 47}	&  0.0393 &  20   &   123     &       {\bf 49}        &  0.4318 &  -15  &   160     \\
{\bf 3 }	&  0.0446 &  14   &   130     &       {\bf 10}        &  0.4487 &  6	&   139     \\
{\bf 9 }	&  0.0744 &  -22  &   168     &       {\bf 18}        &  0.4979 &  67	&   73      \\
{\bf 17}	&  0.1369 &  -84  &   235     &       {\bf 26}        &  0.4993 &  69	&   71      \\
{\bf 51}	&  0.1497 &  -95  &   246     &       {\bf 27}        &  0.5022 &  73	&   67      \\
{\bf 25}	&  0.1705 &  -109 &   261     &       {\bf 4 }	      &  0.5103 &  83	&   56      \\
{\bf 12}	&  0.1739 &  -111 &   263     &       {\bf 13}        &  0.5260 &  103  &   35      \\
{\bf 1 }	&  0.2184 &  -130 &   283     &       {\bf 5 }	      &  0.5448 &  127  &   10      \\
{\bf 28}	&  0.2302 &  -132 &   286     &       {\bf 20}        &  0.5724 &  160  &   -26     \\
{\bf 22}	&  0.2725 &  -132 &   285     &       {\bf 30}        &  0.5969 &  187  &   -55     \\
{\bf 6 }	&  0.2777 &  -131 &   284     &       {\bf 14}        &  0.6189 &  208  &   -78     \\
{\bf 29}	&  0.2888 &  -128 &   281     &       {\bf 21}        &  0.6235 &  213  &   -82     \\
{\bf 24}	&  0.2890 &  -128 &   281     &       {\bf 23}        &  0.6545 &  238  &   -110    \\
{\bf 44}	&  0.3017 &  -123 &   276     &       {\bf 15}        &  0.6608 &  243  &   -114    \\
{\bf 38}	&  0.3087 &  -120 &   273     &       {\bf 46}        &  0.6749 &  251  &   -124    \\
{\bf 2 }	&  0.3119 &  -118 &   271     &       {\bf 8 }	      &  0.6907 &  260  &   -133    \\
{\bf 39}	&  0.3267 &  -110 &   263     &       {\bf 50}        &  0.7704 &  272  &   -146    \\
{\bf 31}	&  0.3333 &  -106 &   258     &       {\bf 16}        &  0.7781 &  270  &   -144    \\
{\bf 45}	&  0.3338 &  -106 &   258     &       {\bf 11}        &  0.8155 &  257  &   -129    \\
{\bf 7 }	&  0.3350 &  -105 &   257     &       {\bf 19}        &  0.8567 &  230  &   -100    \\
{\bf 40}	&  0.3387 &  -103 &   255     &       {\bf 52}        &  0.8900 &  200  &   -69     \\
{\bf 41}	&  0.3511 &  -94  &   245     &       {\bf 33}        &  0.9414 &  143  &   -8      \\
{\bf 42}	&  0.3685 &  -80  &   230     &       {\bf 34}        &  0.9601 &  120  &   16      \\
{\bf 43}	&  0.3740 &  -75  &   225     &       {\bf 35}        &  0.9792 &  96	&   42      \\

\noalign{\smallskip}
\hline
\end{tabular}}
\end{table}

\begin{table}[h]
\caption{Stellar parameters of the non-eclipsing binary (HD\,64315\,AaAb) derived from the  
radial velocity curves. \label{t5}}    
\centering 
\scalebox{0.85}{   
\begin{tabular}{l c c }        
\hline\hline
\noalign{\smallskip}           
&Aa& Ab\\
\noalign{\smallskip}  
\hline
\noalign{\smallskip}  
Orbital period (day)&\multicolumn{2}{c}{2.70962901 $\pm$ 0.00000021}\\
Zero point of ephemeris (HJD)&\multicolumn{2}{c}{2454022.12350 $\pm$ 0.00012}\\
Eccentricity&\multicolumn{2}{c}{0 (assumed)}\\
Longitude of periastron ($^{\circ}$)&90&270\\
Systemic velocity (\kms)&\multicolumn{2}{c}{69.98 $\pm$ 0.03}\\
Semi-amplitude of velocity (\kms)&203.74$\pm$0.06&217.52 $\pm$ 0.06\\
Projected semimajor axis (\Rsun)&10.907 $\pm$ 0.003&11.644 $\pm$ 0.003\\
Minimun mass (\Msun)&10.838 $\pm$ 0.005&10.151$\pm$ 0.005\\
Mass ratio ($M_{2}/M_{1}$)&\multicolumn{2}{c}{0.9366 $\pm$ 0.0006}\\
\hline 
\noalign{\smallskip}  
\end{tabular}}
\end{table}

\subsubsection{Eclipsing binary (BaBb)} \label{EB}

\begin{figure}[h]
\centering
\includegraphics[width=8 cm,angle=0]
{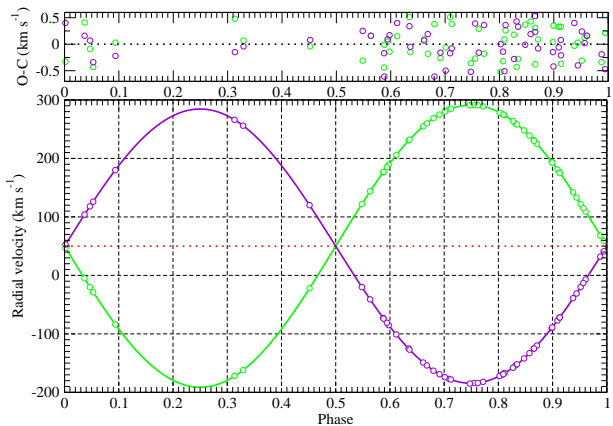}
\caption{Radial velocity curves fitted to the observational data and shown against orbital phase 
(blue line: Ba; red line: Bb).
The red dotted line corresponds to the systemic velocity.
The residuals are shown in the top panel.}
\label{f8}
\end{figure}

We derived the radial velocity curve (see Figure~\ref{f8})  and the corresponding orbital parameters, by
using SBOP and the same methodology as in Sect.~\ref{nEB}. The mass ratio, semi-major axis and systemic
velocities obtained were included as guess parameters in the generalised Wilson-Devinney (WD) code \citep{wils1971} 
in its 2010 version. The period and zero point ephemeris derived from the spectroscopic data are $1.018965\:$d and 
HJD 2452550.6272, respectively. Radial velocities ordered according to the phase are shown in Table~\ref{t3cd}. 
The tiny residuals are a consequence of the same procedure discussed in Sect.~\ref{nEB}. 

\begin{table}[h]
\caption{Radial velocities in phase of the eclipsing binary (B). \label{t3cd}}
\centering 
\scalebox{0.94}{
\begin{tabular}{c c c c c c c c}
\hline \hline
\noalign{\smallskip}
$\sharp$&phase &RV$_{Ba}$&RV$_{Bb}$&$\sharp$&phase &RV$_{Ba}$&RV$_{Bb}$\\
\noalign{\smallskip}
\hline
\noalign{\smallskip}  
{\bf 27 }     &  0.0017  &  47   &   53   &   {\bf 43}      &  0.7619  &  291  &   -184 \\
{\bf 45 }     &  0.0368  &  -5   &   104  &   {\bf 36}      &  0.7715  &  289  &   -182 \\
{\bf 25 }     &  0.0467  &  -20  &   118  &   {\bf 11}      &  0.8022  &  278  &   -172 \\ 
{\bf 1  }     &  0.0527  &  -29  &   126  &   {\bf 4 }	    &  0.8065  &  276  &   -170 \\ 
{\bf 29 }     &  0.0936  &  -84  &   180  &   {\bf 37}      &  0.8091  &  275  &   -169 \\ 
{\bf 22 }     &  0.3134  &  -172 &   266  &   {\bf 8 }	    &  0.8112  &  274  &   -167 \\ 
{\bf 23 }     &  0.3292  &  -162 &   256  &   {\bf 10}      &  0.8268  &  264  &   -158 \\ 
{\bf 20 }     &  0.4518  &  -22  &   120  &   {\bf 9 }	    &  0.8315  &  260  &   -154 \\ 
{\bf 19 }     &  0.5484  &  122  &   -20  &   {\bf 52}      &  0.8343  &  258  &   -152 \\ 
{\bf 24 }     &  0.5635  &  144  &   -41  &   {\bf 6 }	    &  0.8473  &  248  &   -142 \\ 
{\bf 21 }     &  0.5877  &  176  &   -73  &   {\bf 50}      &  0.8573  &  239  &   -133 \\ 
{\bf 38 }     &  0.5881  &  177  &   -74  &   {\bf 7 }	    &  0.8653  &  231  &   -125 \\ 
{\bf 18 }     &  0.5944  &  185  &   -81  &   {\bf 51}      &  0.8656  &  230  &   -125 \\ 
{\bf 14 }     &  0.5977  &  189  &   -85  &   {\bf 31}      &  0.8710  &  225  &   -120 \\ 
{\bf 33 }     &  0.6117  &  206  &   -101 &   {\bf 5 }	    &  0.8983  &  194  &   -90  \\ 
{\bf 17 }     &  0.6344  &  231  &   -125 &   {\bf 48}      &  0.8993  &  193  &   -89  \\ 
{\bf 39 }     &  0.6360  &  232  &   -127 &   {\bf 3 }	    &  0.9089  &  181  &   -77  \\ 
{\bf 34 }     &  0.6613  &  255  &   -149 &   {\bf 30}      &  0.9128  &  176  &   -72  \\ 
{\bf 40 }     &  0.6681  &  260  &   -154 &   {\bf 47}      &  0.9130  &  175  &   -72  \\ 
{\bf 16 }     &  0.6803  &  269  &   -163 &   {\bf 28}      &  0.9377  &  142  &   -39  \\ 
{\bf 13 }     &  0.6914  &  275  &   -169 &   {\bf 46}      &  0.9441  &  133  &   -31  \\ 
{\bf 41 }     &  0.7010  &  280  &   -174 &   {\bf 44}      &  0.9515  &  122  &   -20  \\ 
{\bf 15 }     &  0.7091  &  284  &   -177 &   {\bf 49}      &  0.9566  &  115  &   -13  \\ 
{\bf 35 }     &  0.7122  &  285  &   -178 &   {\bf 2 }	    &  0.9608  &  109  &   -7	\\ 
{\bf 42 }     &  0.7473  &  291  &   -185 &   {\bf 32}       &  0.9879  &  68	&   32   \\
{\bf 12 }     &  0.7551  &  291  &   -184 &   {\bf 26}       &  0.9942  &  59	&   41   \\

\noalign{\smallskip}
\hline
\end{tabular}}
\end{table}

We obtained the light curve and radial velocity curve models by computing the parameters via 
differential corrections until all free parameter adjustment of 
light curves and radial velocity curves is reached according to the least-squares criterion. 
We assumed that stars Ba and Bb form a contact system, in which case circularisation and synchronisation 
are acceptable approximations. We chose mode 1 of the WD code, corresponding to overcontact binaries. In this case, 
the surface potentials are the same for both stars ($\Omega_{1}=\Omega_{2}$). The radiative model for both components 
of the binary system is an atmosphere model by \cite{kuru1993}. The surface is divided into a grid of $40 \times 40$ 
elements for each star. To improve the convergence of the solution, 
we chose symmetrical derivatives \citep{wils1976b}. The code can apply the detailed reflection model of \cite{wils1990}, 
a treatment especially recommended with overcontact binaries. We have also considered proximity effects on both stars. 
A square root limb-darkening law was applied during the process, as it is an order of magnitude more precise than the 
linear law \citep{hamm1993}. The bolometric albedos of both components were fixed at $A_1=A_2=1$, because the
atmospheres are expected to be in radiative equilibrium \citep{vonz1924}. Because of
local energy conservation, this also implies gravity brightening exponents $g_1=g_2=1$. Other constraints applied 
are described in mode 1 of the WD code. Both temperatures were fixed to 32\,000~K, in agreement with the expectation of a spectral type
not later than O9.5\,V nor earlier than O8.5\,V (as both components have to be O-type stars, and at the same time considerably fainter than the $\sim$O6\,V components that dominate the spectrum). The temperature ratio is not an adjustable parameter during the convergence process. 
This is a compulsory constraint due to the morphology of the binary star, where both components share a volume 
of their Roche lobes and thermal contact is assumed. The third light is included due to the presence of the binary system A.

The process to convergence of all free parameters is iterative and simultaneous for all observables analysed, i.e. the radial velocity curves and light curves. The criterion for convergence adopted is as follows. For three consecutive iterations, all adjustable parameters must be within two standard deviations. Once convergence is reached, five solutions are derived by varying the parameters within the standard deviation and fitting the observations again. We choose the fit with the smallest dispersion as a final solution.

\begin{figure}[h]
\centering
\includegraphics[width=8 cm,angle=0,clip]
{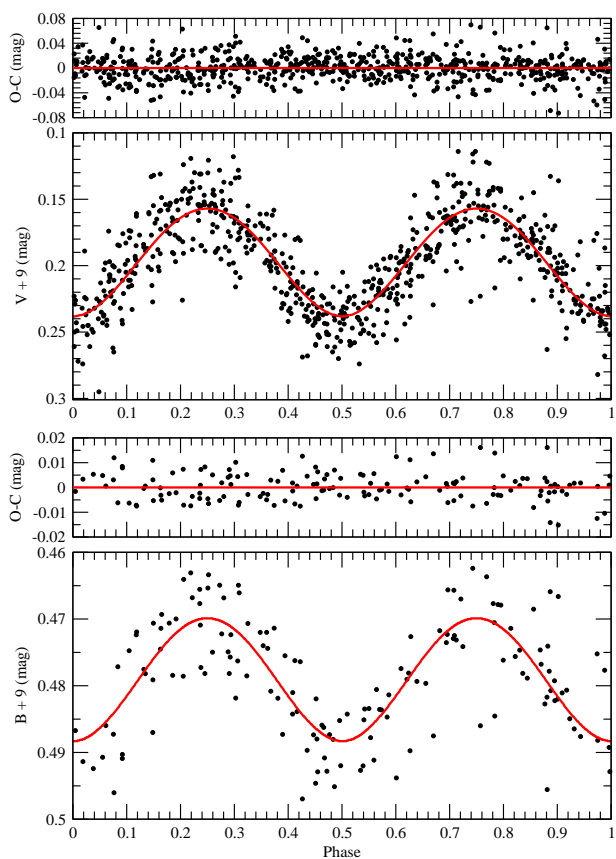}
\caption{Light curve model fitted to the observational data and residuals in $V$ filter (above) and $B$ filter (below).}
\label{f9}
\end{figure}

Light curve models are shown in Figure\ref{f9}. Their shape phenomenologically corresponds to an eclipsing variable of the EW type; there is no plateau between the eclipses. The light curve exhibits a continuous and monotone 
shape along the cycle. This shape of the model light curve confirms that the two stars are 
overfilling and sharing their Roche lobes. Light curves show a significant dispersion. As a consequence, it
is difficult to visually distinguish which one of the two minima is deeper. For a best estimate of the difference in depth, we averaged all photometric 
data points between $\phi=0.99$ and~$0.01$, and all points between $\phi=0.49$ 
and~$0.51$, obtaining a difference of 5 millimagnitudes for the $V$ filter and 2 millimagnitudes for the $B$ filter. The latter value is not significant because of the small number of points. Both differences are smaller than the intrinsic dispersion ($\sigma_V=0.02$; $\sigma_B=0.006$) of the corresponding light curves, and so we have to conclude that both minima are of equal depth. Residuals 
are under 0.08~mag for the $V$ filter and under 0.02~mag for the $B$ filter, showing a reasonably good fit in either case,
in spite of the low quality of the photometric data. 

The linear ephemeris equation, where the epoch of successive times of primary-eclipse minima (phase zero), 
$T_{{\rm min}}$, is calculated from the period and zero-time ephemeris derived from the light curves, is

\begin{multline*}
 T_{\rm min} = {\rm HJD}~(2452550{.}62838\pm0{.}00018)\\
  + (1^{\rm d}{.}0189569\pm0^{\rm d}{.}0000008)\times E \:\,
\end{multline*}

All the stellar and orbital parameters are shown in Table~\ref{t6}.
The difference between the spectroscopic and photometric period is about seven-tenths of one second, and the zero-time ephemeris
derived from spectroscopic and photometric data differ by about 100~s, demonstrating excellent agreement, 
given the complex methodology and low quality of the data. The systemic velocity transformed to the LSR 
coordinates would be $+35\:$\kms, significantly different from the systemic velocity of the non-eclipsing binary (AaAb). 
Both components have
the same mass, $14.6\:$\Msun. In view of this, we assume that both components have spectral type O9.5\,V. As is typical in this kind of early-type overcontact system, the surface effective gravity obtained is quite high, 
typical of zero-age main-sequence (ZAMS) stars.

\begin{table}[h]
\caption{Stellar parameters for the eclipsing binary (HD\,64315\,BaBb) derived from the combined analysis of the 
radial velocity curves and photometric light curves. \label{t6}}    
\centering 
\scalebox{0.85}{   
\begin{tabular}{l c c }        
\hline\hline
\noalign{\smallskip}           
&Ba& Bb\\
\noalign{\smallskip}  
\hline
\noalign{\smallskip}  
Orbital period (day)&\multicolumn{2}{c}{1.0189569 $\pm$ 0.0000008}\\
Zero point of ephemeris (HJD)&\multicolumn{2}{c}{2452550.62838 $\pm$ 0.0018}\\
Eccentricity&\multicolumn{2}{c}{0 (assumed)}\\
Inclination ($^{\circ}$)&\multicolumn{2}{c}{48.2 $\pm$ 1.4}\\
Longitude of periastron ($^{\circ}$)&90&270\\
Systemic velocity (\kms)&\multicolumn{2}{c}{50.4 $\pm$ 9.2}\\
Semi-amplitude of velocity (\kms)&243.0$\pm$15.7&243.2 $\pm$ 15.7\\
Semi-major axis (\Rsun)& \multicolumn{2}{c}{13.1 $\pm$ 0.7}\\
Surface normalised potential&\multicolumn{2}{c}{3.58 $\pm$ 0.10}\\
Mass (\Msun)&14.6 $\pm$ 2.3&14.6$\pm$ 2.3\\
Mass ratio ($M_{2}/M_{1}$)&\multicolumn{2}{c}{1.00 $\pm$ 0.06}\\
Mean equatorial radius (\Rsun)&5.52 $\pm$ 0.55&5.33 $\pm$ 0.52\\
Polar radius (\Rsun)&4.96 $\pm$ 0.47&4.82 $\pm$ 0.45\\
Side radius (\Rsun)&5.26 $\pm$ 0.55&5.08 $\pm$ 0.52\\
Back radius (\Rsun)&5.84 $\pm$ 0.77&5.57 $\pm$ 0.69\\
Projected rotational velocity\tablefootmark{a} (\kms)&203 $\pm$ 15&198 $\pm$ 14\\
Surface effective gravity\tablefootmark{b} ($\log\,g$) &4.19 $\pm$ 0.05&4.16 $\pm$ 0.05\\
Luminosity ratio (V-filter)($L_{2}/L_{1}$)&\multicolumn{2}{c}{0.897 $\pm$ 0.023}\\
Luminosity ratio (B-filter)($L_{2}/L_{1}$)&\multicolumn{2}{c}{0.889 $\pm$ 0.033}\\
Third light (V-filter) ($l_{3}$)&\multicolumn{2}{c}{0.462 $\pm$ 0.044}\\
Third light (B-filter) ($l_{3}$)&\multicolumn{2}{c}{0.579 $\pm$ 0.018}\\
\hline 
\noalign{\smallskip}  
\end{tabular}}
 \tablefoot{\scriptsize{
 \tablefoottext{a}{calculated from the mean equatorial radius};
 \tablefoottext{b}{calculated from the side radius}}}
\end{table}

\begin{figure}[h]
\centering
\includegraphics[width=8 cm,angle=0]
{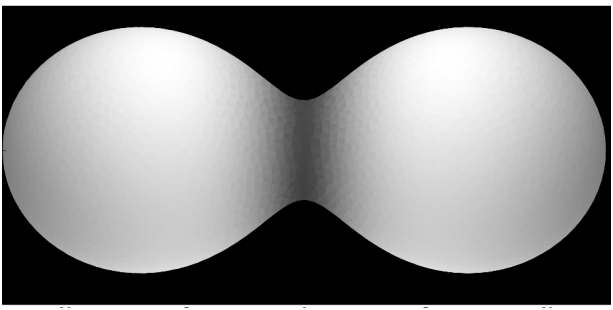}
\caption{Representative drawing of HD\,64315\,BaBb to scale at quadrature phase, created with the {\sc phoebe} 
2.0-alpha code via the Python interface.}
\label{i3}
\end{figure}

The radii derived from the surface potential show that the two stars are sharing their atmosphere.
 We provide a representative
drawing (see Figure~\ref{i3}) where we can see how the atmosphere of the two stars overlap.

\section{Direct distance estimation} \label{ded}

Eclipsing binaries allow the derivation of geometrical distances to the systems, which can be very precise 
\citep[e.g.][]{sout2004, vila2010}. In overcontact binaries, there are many complications, owing to the 
interaction and geometrical distortion, but a direct distance estimate is still possible from the stellar 
parameters. In HD~64315, we find the added complication of a third body (the detached non-eclipsing binary) 
that is brighter than the eclipsing binary.
For the calculation, we assumed effective temperatures of 32\,000~K for components Ba and Bb, and 40\,000~K for components
Aa and Ab. This is an approximation based on qualitative criteria described in Sects.~\ref{nEB} and \ref{EB}
and the SpT-\Teff~calibrations by \cite{mart2005}. 
The apparent magnitudes of every star
(for every passband) were calculated and included in Table~\ref{t7}. We assumed the uncertainties 
to be twice the standard deviation of the corresponding light curve. 
Then, we estimated a distance-dependent flux, in standard physical units, 
including those proximity effects supported by the model and accepting constraints relative to the
radii of components of Aa and Ab ($R_{\textrm{Aa}}^{\textrm{lobe}}=11.5\:$\Rsun\, and $R_{\textrm{Ab}}^{\textrm{lobe}}=11.2\:$\Rsun). This procedure, although approximate, is more 
accurate than the simple use of the mean radii. We follow the procedure described in \cite{vila2010}.
The parameters used  are shown in Table~\ref{t7}, together with our estimates of distance, $4.7\pm0.6$~kpc,
for the non-eclipsing binary (AaAb) and $5.0\pm0.8$~kpc for the eclipsing binary (BaBb). Both values are fully consistent. 
The errors are mainly due to the relatively large uncertainties in the extinction and the bolometric 
correction.

\begin{table*}
\caption{Parameters used to estimate the distance to HD\,64315. \label{t7}}    
\centering 
\scalebox{0.9}{   
\begin{tabular}{l c c c c}  
\hline\hline
Binary properties&\multicolumn{2}{c}{non-eclipsing binary (A)}&\multicolumn{2}{c}{eclipsing binary (B)}\\
\noalign{\smallskip} 
\hline
\noalign{\smallskip} 
$m_V$ (mag)    & \multicolumn{2}{c}{$9.32\pm0.04$}&\multicolumn{2}{c}{$11.31\pm0.04$}\\
$m_B$ (mag)    & \multicolumn{2}{c}{$9.63\pm0.012$}&\multicolumn{2}{c}{$11.62\pm0.012$}\\
$E(B-V)$ (mag)  & \multicolumn{2}{c}{$0.62\pm0.04$}&\multicolumn{2}{c}{$0.60\pm0.04$}\\
$A_V$           & \multicolumn{2}{c}{$1.93\pm0.23$}&\multicolumn{2}{c}{$1.86\pm0.22$}\\
$M_V$ (mag)     & \multicolumn{2}{c}{$-5.97\pm0.18$}&\multicolumn{2}{c}{$-4.04\pm0.26$}\\
$(V_0-M_V)$ (mag)&\multicolumn{2}{c}{$13.36\pm0.29$}&\multicolumn{2}{c}{$13.5\pm0.3$}\\
distance (pc)    &\multicolumn{2}{c}{$4700\pm600$}&\multicolumn{2}{c}{$5000\pm800$}\\
\noalign{\smallskip} 
\hline
\noalign{\smallskip} 
Component properties& star Aa&star Ab&star Ba&star Bb\\
\noalign{\smallskip} 
\hline
\noalign{\smallskip} 
$\log\,(L$/ \Lsun)&$5.46\pm0.08$&$5.43\pm0.08$&$4.46\pm0.11$&$4.42\pm0.11$\\
$ M_V$    (mag) &$-5.25\pm0.18$&$-5.19\pm0.18$&$-3.34\pm0.26$&$-3.23\pm0.26$\\
$(B-V)_0$ (mag) &$-0.311\pm0.005$&$-0.311\pm0.005$&$-0.290\pm0.006$&$-0.290\pm0.007$\\
\noalign{\smallskip} 
\hline 
\end{tabular}}
\end{table*}

% ------------------------------------------------------------------------
\section{Discussion}\label{discusion}
\begin{figure}[h]
\centering
\includegraphics[width=7.8 cm,angle=0]
{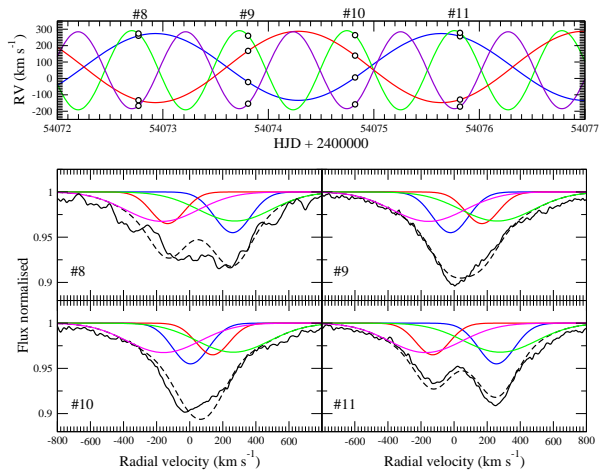}
\caption{Radial velocity curves of the four stars of HD\,64315 in time (above; Aa: blue,
Ab: red, Ba: green, Bb: magenta). The \ioni{He}{II} line of each spectrum is shown (solid black line) compared to the model 
from all components (dashed line).}
\label{f11}
\end{figure}

% ------------------------------------------------------------------------
HD~64315 presents evident difficulties in the determination of its orbital and stellar parameters. Firstly,
all components have broad and shallow spectral lines, because of their high rotational velocities. As a consequence, 
cross-correlation techniques are difficult to apply because the peaks of the spectral lines for
every component are not distinguishable, thereby increasing the uncertainties. Moreover, the short 
periods of both binaries add further complication to any disentangling process. As an example, in Figure~\ref{f11} 
we show radial velocity curves for all components as a function of time (HJD\,=\,2454072$-$2454077). 
Spectra 8, 9, 10 and~11 are marked on the radial velocity curve with their corresponding times. 
The spectral lines of \ioni{He}{ii} corresponding to these four spectra are displayed together with the contribution 
of every component and the sum of these Gaussian functions. Apparently, radial velocities in spectra \#8 and \#11 
are very similar
quantitatively, but the morphology of the spectral lines is totally different: while \#11 shows two peaks,
\#8 shows an erratic profile aggravated by its low S/N. In the case of spectra \#9 and \#10, we
note the asymmetries of the spectral lines, and how a spectral line showing a single peak can hide four lines coming from the four components. The contributions of Ba and Bb to the profile are shallow and weak, and, given the higher orbital velocity of system B, on many occasions appear as very faint extended wings on the sides of the lines due to system A or, when their radial velocities are low, are completely lost inside the lines from A. 

Despite these difficulties, our methodology is able to derive orbital and stellar parameters with a very limited set of assumptions. We find that HD~64315 is a quadruple star, consisting of two binary systems separated by about 100\,000\,\Rsun\,
(455 A.U.) at a distance of around $5\:$kpc. The non-eclipsing system (O6\,V+O6\,V) is a detached system, 
However, the inclination derived suggests that the two components are not far from filling their 
respective Roche lobes, and we have used this approximation for some estimates.

The eclipsing binary (O9.5\,V+O9.5\,V) is a contact system with a very short period. We consider this object a 
very strong merger progenitor candidate. To estimate when the merger will take place, we assume that the coalescence 
of the two components will occur when the outer Roche lobe radius is reached. Application of the equation derived by 
\cite{yaku2005} leads to an outer Roche lobe radius for HD\,64315\,Ba of $6.3\:R_{\sun}$. If we assume that
the Kevin-Helmholtz time is the merging time scale, the duration of the merger will be 26\,000 years. The two components 
of HD\,64315\,B are already overfilling the volume of their  Roche lobe, presenting an overlapping volume between them. 
If we take, for comparison a Geneva track \citep{geor2013} for a $15\:M_{\sun}$ with $Z=0.01$ and 
$\Omega$/$\Omega_{\textrm{crit}}=0.568$ (equivalent to $v$/$v_{\textrm{crit}}$ =0.4), we see that a polar radius 
$\approx5\:R_{\sun}$ (as we find from our solution) corresponds to an age around $3\:$Myr (in good accord with 
the presence of an \ion{H}{ii} region around the star). If the star was isolated and stellar evolution was the only 
driver of shape changes, an equatorial radius $\approx 6.3\:R_{\sun}$ would be reached by an age of $\la 8\:$Myr. 
However, in the case of HD\,64315\,Ba, we see that the distortion from a spherical shape is much larger than caused by 
rotation alone. Alternatively, a decrease of 2\,000~K in its $T_{\textrm{eff}}$ would also imply overfilling the outer Roche lobe. 
Again, the evolutionary timescale for an isolated star to achieve that stage is a few Myr.

In the case of HD\,64315\,Aa, the outer Roche lobe radius of HD\,64315\,Aa is $15.1\:R_{\sun}$, and so the stars are very far 
away from the merging condition, as expected, given the much longer period. For two stars of this size, the merging 
timescale would be around 10\,000 years. In both cases, the merger happens in about of the lifetime of an individual
star with the same spectral type.

The two spectroscopic binaries that we have observed must correspond to the two objects separated by speckle interferometry, 
as we do not see any stationary component in the spectra. There may be further components, but then they have to 
be fainter than binary B. With a separation of $\approx455\:$AU, 
the two systems must orbit each other with a period of $\sim1000\:$years. They are thus likely gravitationally bound.

\subsection{Astrophysical context}

It is difficult to find stellar objects with similar characteristics in the literature, as most of the 
quadruple system studied correspond to visual binaries or systems with much lower masses. A notable example is the quadruple system QZ~Car \citep{mayer01}, composed of binaries A (O9.7\,I+b2\,v, $\,P_{\textrm{A}}=21\:$d) and 
B (O8\,III+o9\,v,$\,P_{\textrm{B}} = 6\:$d). It presents several differences with HD~64315: the primaries in the two binaries within QZ~Car are evolved stars, and the secondaries are rather less massive. Moreover, the common orbit has a period likely measured in decades rather than centuries.

During the last decade, several studies have found the multiple nature of systems previously believed to be binaries, for example, LY Aur \citep{maye2013} or SZ~Cam \citep{tama2012}, the latter in the central dense region of the open cluster NGC~1502.  We expect that many binary systems known to present spectral asymmetries without explanation or deviating radial velocities will be found to be multiple in the near future \citep[cf.][]{rapp2013, lohr2015}, probably resulting in a need to review statistics about percentages of binary and multiple stars in young clusters. 

The two binaries within HD~64315, on the other hand, present quite typical characteristics. A system very similar  to HD\,64315\,A is DH~Cep. It has a very similar orbital period of 2.8~d. The components are classified as O5.5\,V and O6.5\,V, and their masses, as derived by \citet{hild1996}, are around $33\:M_{\sun}$ for the primary and $30\:M_{\sun}$ for the secondary. As in  HD\,64315\,A, DH~Cep is not an eclipsing binary and the measured rotational velocities are around $150\:$\kms. Since DH~Cep is not complicated by the presence of a second binary, \cite{hild1996} can derive accurate radii, finding values $\approx8.5\:$\Rsun\ for both components, well within their Roche lobes.

HD\,64315\,B is an overcontact system, and one of the most massive ones known. Among systems containing O-type stars, only OGLE SMC-SC10 108086 has a shorter orbital period of 0.88~d \citep{hild05}. With masses of 14 and $17\:$\Msun, this SMC system is quite similar to HD\,64315\,B. The more massive LMC system VFTS 352 has an orbital period of 1.12~d \citep{almeida15}. Another massive binary with a shorter period, GU~Mon, contains later-type components (B1\,V+B1\,V) with lower masses of around $9\:$\Msun\ \citep{lore2016}. All these objects have been considered likely merger binaries. HD\,64315\,B is exceptional among them, because of its membership in a multiple system. Its eventual merger will lead to the formation of a hierarchical triple system where all the components have about equal masses.

\subsection{Formation and environment}

We present different estimates of the distance to HD~64315, all agreeing on $\approx5$\:kpc. Previous studies had not considered the multiple nature of the system and had obtained distances $\la4\:$kpc, based on calibrations of the luminosity for the spectral type assumed. From the interstellar lines seen in our spectra,
we have identified a local component with a kinematic distance of $\sim5\:$~kpc. The direct distance estimation has also given distances of $4.7\pm0.6$~kpc and $5.0\pm0.8$~kpc for the two binary systems. All measurements are clearly compatible. This new value brings HD~64315 to the same distance as most recent estimates for Haffner~18 \citep[and references therein]{yadav15}. However, several recent papers point to a rather higher distance for Haffner~19, which is visually closer to HD~64315 (only $4\farcm5$ away) and, unlike Haffner~18, contains an O-type star. We must look with some scepticism to these long distance determinations for Haffner~19, not only because of the morphological reasons presented by \citet{sn09}, but also because the ionised gas around the cluster has essentially the same radial velocity as that of Sh2-311. It would be extremely surprising to see a distant very young cluster without any associated nebulosity through the molecular cloud associated with Sh2-311, especially if we consider the low reddening, $E(B-V)\approx0.6$~mag, to Haffner~19. Pending accurate \textit{Gaia} distances, we advance that Sh2-311, at a Galactocentric distance of $\sim12\:$kpc, has a subsolar composition, in agreement with the Galactic abundance gradient \citep{rojas05,rodri10} and thus the use of solar abundance tracks may result in overestimated distances.

HD~64315 lies in isolation near the centre of Sh2-311. Observations by \citet{yadav15} show that most of the bright stars in its surroundings are foreground late-B objects. Indeed, using photometry of a circular area of radius $\sim3\arcmin$ around HD~64315 they only find two objects with colours compatible with a young population associated w the \ion{H}{ii} region until they reach $\sim4\:$mag fainter. Given our total $M_{V}\approx-6$, this means that there are at most two O-type or early-B stars (earlier than $\sim$B2) in the immediate vicinity of HD~64315. This is a very unusual configuration for a star sitting in the middle of an \ion{H}{ii} region with active star formation \citep{sn09}, since generally the mass of the most massive star in a cluster correlates with cluster mass  \citep{weidner2010}. Against a deterministic interpretation of this correlation, \citet{oey2013} presented a sample of 14 OB stars in the SMC that meet strong criteria for having formed under extremely sparse star-forming conditions in the field. HD~64315 can be interpreted as a Galactic equivalent to these objects, However, it also represents a cautionary tale about the meaning of an `isolated' star. 

Was this complex multiple system born in isolation in the middle of Sh2-311? Though unusual, this scenario looks quite likely. The only possible alternative, if Haffner~19 is really at the same distance as HD~64315, against recent analyses, is that HD~64315 is a runaway from this cluster. However, it is extremely difficult to conceive a dynamical interaction that can result in the ejection of such a wide binary without disrupting it. Of course, if the two binaries are not physically bound, the ejection scenario also breaks down, because they should have been ejected individually in exactly the same direction with exactly the same velocity, an even more improbable occurrence. Finally, the systemic radial velocity of binary A, which contains most of the system mass, is quite similar to that of the surrounding medium, again suggesting in situ formation.

We cannot exclude the possibility that HD~64315 is really a compact cluster; most of the mass is concentrated in the two observed binaries. Lower-mass components would not be observable in the glare of this very bright system. However, in any case, it is highly unlikely that HD~64315 is surrounded by a cluster with $\sim1\,000\:M_{\sun}$, as is usual for other O6\,V stars \citep{weidner2010}. 

Binary separation in multiple stars is a possible indicator to discern between two of the main mechanisms proposed for stellar formation. Turbulent fragmentation leads to initial separations $>\,$500 AU
\citep{offn2010}, while disk fragmentation produces initial separations $<\,$500 AU. Unfortunately, the separation between the binaries A and B is around 500 AU, just between the predictions of the two theories. Another criterion used to distinguish between these two formation scenarios is the
alignment of the stellar spin \citep{offn2016}. When formed via disk fragmentation, stars have common angular momenta and therefore aligned stellar spin. The fact that the inclinations of the two binaries are quite similar supports in this case the disk fragmentation model, even though this argument has no statistical significance on its own.

% ------------------------------------------------------------------------
\section{Summary and conclusions}\label{summary}
% ------------------------------------------------------------------------

By using a complex procedure to analyse 52 high-resolution spectra, we are able to confirm that HD~64315 
contains two binary systems, one of which is an eclipsing binary. The non-eclipsing binary (system A) has a period of 2.7~d, 
and is quite similar to the well-studied DH~Cep. Its components are hotter and more luminous than those of the eclipsing binary 
(system B), and dominate the appearance of the system. System A is a detached binary composed of two stars with spectral types 
around O6\,V, with minimum masses of $10.8\:$\Msun\ and $10.2\:$\Msun, and likely masses $\approx30\:$\Msun. The eclipsing binary 
has a shorter period of 1.0~d, and produces a weak, but observable effect in the system light curve. We have derived masses of 
$14.6 \pm 2.3\:$\Msun\ for both components, which are late O-type stars. They are almost identical: they overfill their 
respective Roche lobes and share a common envelope. System B is thus one of the most massive overcontact binaries known, and 
a very likely merger progenitor. Its merger within such a complex system may lead to the formation of a hierarchical triple 
system with three stars of similar masses.

We are not able to rule out an accompanying low-mass compact cluster with current observations, but 
HD\,64315 has few nearby OB-type companions and does not appear to have been ejected from a nearby 
open cluster. It thus seems likely that the system, with a total mass above $90\:$\Msun, formed in relative 
isolation near the centre of the Sh2-311 \ion{H}{ii} region. In summary, HD\,64315 is potentially a massive hierarchical 
system that formed in a sparse environment, which nicely highlights the need for detailed studies of multiplicity 
in apparently `isolated' stars.

% ------------------------------------------------------------------------
\begin{acknowledgements}
This research is partially supported by the Spanish Government Ministerio de Econom\'{i}a y Competitivad 
(MINECO/FEDER) under grants AYA2015-68012-C1-1-P and AYA2015-68012-C2-2-P. This research has made use of the Simbad, Vizier 
and Aladin services developed at the Centre de Donn\'ees Astronomiques de Strasbourg, France. 
Francesc Villardel acknowledges the support of the Spanish Ministry for Economy and Competitiveness (MINECO) 
and the Fondo Europeo de Desarrollo Regional (FEDER) through grant ESP2016-80435-C2-1-R, as well 
as the support of the Generalitat de Catalunya/CERCA programme.

\end{acknowledgements}
% ------------------------------------------------------------------------

%
% ------------------------------------------------------------------------

% ------------------------------------------------------------------------
%
\end{document}